\documentclass[aps,prd,notitlepage,nofootinbib,superscriptaddress, twocolumn]{revtex4}
\usepackage{amssymb}
\usepackage{color}
\usepackage{amsmath,amsxtra}
\usepackage{verbatim}
\usepackage{appendix}
\usepackage[pdftex]{graphicx}
\usepackage[usenames,dvipsnames]{xcolor}
\usepackage{amsfonts}
\usepackage{wasysym}

\begin{document}

\title{Inflationary spectral tilts as a result of the dilatation symmetry breaking}

\author{Pisin Chen}
\affiliation{Leung Center for Cosmology and Particle Astrophysics, National Taiwan University, Taipei, Taiwan 10617}
\affiliation{Department of Physics and Center for Theoretical Sciences, National Taiwan University, Taipei, Taiwan 10617}
\affiliation{Kavli Institute for Particle Astrophysics and Cosmology, SLAC National Accelerator Laboratory, Stanford University, Stanford, CA 94305, USA}
\author{Jiro Matsumoto}
\author{Rio Saitou}
\email{rsaito@ntu.edu.tw}
\affiliation{Leung Center for Cosmology and Particle Astrophysics, National Taiwan University, Taipei, Taiwan 10617}

%abstruct===========================================================
\begin{abstract}

We derive the spectral indices and their runnings of single inflation models by a new approach. We perform a dilatation transformation to the linear cosmological perturbations and derive a current (non-)conservation law. Using it, we construct a dilatation charge and a Ward-Takahashi identity
for the two-point correlators, 
and derive two \textit{exact} expressions for the tree-level spectral indices.
First, we apply the slow roll expansion to one of the exact expressions. We calculate the spectral indices and their runnings up to the second and the third order of slow roll parameters respectively, with use of the ``horizon crossing formalism". 
By construction, our results are more rigorous and generic than the previous works. Then, we analyze another exact expression to understand how the perturbations and the slow roll parameters contribute to the spectral indices. By a numerical calculation, we confirm that only the behaviors of the slow roll parameters during a few e-folds around the horizon crossing affect
significantly to the values of spectral indices. The analysis in this article indicates that if one cannot use the slow roll parameters, regardless of their values, as the expansion parameters around the horizon crossing, then one can no longer apply the slow roll expansion to the spectral indices, and it is thus necessary to apply the more generic method introduced here. %thus we will need a numerical calculation as here.

\end{abstract}

\maketitle

%Section1==============================================================
\section{Introduction}\label{sec1}

%Introduction of inflation and the requirement for theoretical side
Inflation is the phase of accelerated expansion in the early universe \cite{Guth:1980zm, Starobinsky:1980te, Sato:1980yn}. It can resolve the three big problems of the big bang universe and can create the seed of the large scale structure by the quantum fluctuation around the quasi de Sitter background. Inflation is supposed to end and transit to the big bang universe accompanied with the reheating process. The observation of Cosmic Microwave Background (CMB) anisotropy \cite{Ade:2015lrj, Akrami:2018odb} indicates that the temperature fluctuations of CMB  are essentially gaussian, which is in concordance with the quantum fluctuations predicted by the inflationary theory \cite{Mukhanov:1981xt, Starobinsky:1982ee, Hawking:1982cz, Guth:1982ec}. 
Thus, at the present time, inflation is considered as the standard theory of the early universe.

The observations of CMB anisotropy have imposed significant constraints to the numerous inflationary models. Initiated by the COBE project \cite{Bennett:1996ce}, WMAP \cite{Komatsu:2010fb}, PLANCK \cite{Ade:2015lrj, Akrami:2018odb}, Keck/BICEP \cite{Ade:2018gkx} projects have
clarified that the primordial E-mode power spectrum of the temperature fluctuations is red-tilted in the Fourier space, and there is only a tiny fraction of the B-mode power spectrum. Mainly by the PLANCK, the changing ratio of E-mode spectral tilt and the size of non-gaussianities of E-mode correlations are also constrained. 
While some inflationary models have been excluded by these observational facts, some other models still survive. Not only the potential-driven inflation \cite{Albrecht:1982wi, Linde:1983gd, Steinhardt:1984jj, Martin:2013tda}, but also the kinetically driven inflation \cite{ArmendarizPicon:1999rj, Garriga:1999vw, Kobayashi:2010cm} and their hybrid type models \cite{ArkaniHamed:2003uz, Alishahiha:2004eh, Burrage:2010cu} can still accord with the present observational constraints. To further distinguish and exclude them, we need more precise and accurate observations.
It is expected that the next generation projects, such as PRISM, Lite-BIRD, CMB-S4, AliCPT, would obtain much more precise constraints to the scalar spectral tilt, its running and the ratio of E-mode and B-mode spectra. This, in turn, requires to derive more precise theoretical predictions for those observables.

%present status of theoretical side and its problem
From the theoretical side, the inflationary observables are derived mainly by using the slow roll parameters.
If and only if the slow roll parameters are small enough during the inflation, 
we can use them as expansion parameters for the observables.  Among the observables, the spectral indices and their runnings have been derived up to the second order of the slow roll parameters \cite{Gong:2001he, Habib:2002yi, Martin:2002vn, Casadio:2006wb}. The previous results are, however, restricted to the potential-driven inflation models only. Further, in \cite{Gong:2001he},  the authors imposed an ansatz for the model parameters to derive the power spectrum. In \cite{Martin:2002vn, Casadio:2006wb}, the authors performed an uncertain Taylor expansion for the power spectra, where the convergence of the expansion is not guaranteed. 
Moreover, their predictions rely on the slow roll expansion, so that we cannot apply them to a class of models where the slow roll parameters are not necessarily small throughout the inflation. In fact, many models are reported, in which the predictions of the slow roll expansion crucially deviates from the numerical calculation (e.g. \cite{Starobinsky:1992ts, Leach:2000yw, Adams:2001vc, Makarov:2005uh}). Therefore, the previous methods according to the slow roll expansion are far from rigorous and generic. 

On the other hand, there is another expansion method for the inflationary fluctuations, known as $\delta N$ formalism \cite{Starobinsky:1986fxa, Sasaki:1995aw, Wands:2000dp, Lyth:2004gb}. This method is based on the leading order gradient or momentum expansion parametrized by $\epsilon=k/aH$, where $k$ is the spatial momentum of the fluctuations, $a$ is the scale factor of a local homogeneous universe and $H$ is the Hubble parameter of the local universe. The formalism is reliable up to the first order of $\epsilon$, and it is applicable only for the fluctuations on the super-horizon scales where $\epsilon\ll1$. Thus, we cannot use it for the models in which $O(\epsilon^2)$ terms induce significant effects to the fluctuations. 
Then, the gradient expansion has been recently extended up to the next-to-leading order for limited cases only \cite{Takamizu:2018uty}. 
%Nevertheless, it is not so obvious to express the observables by the model parameters such as the slow roll parameters in the gradient expansion. 
%
Therefore, at the present, we cannot deal strictly with various models which %are out of the applicable range of the theoretical predictions but 
can potentially survive the observational constraints. To face with the coming high precision observations, it is necessary to derive more rigorous and versatile theoretical predictions for the observables. 

%approach and object of this article
In this article, we derive the tree-level spectral indices \textit{exactly} for single field inflation models by a new approach. 
By definition, the indices are intrinsically the scaling dimensions of the two-point correlators in the Fourier space. Hence, if we perform a dilatation transformation to the two-point correlators, we should naturally obtain exact expressions of the spectral indices. 
According to this idea, we define the dilatation transformation to the linear cosmological perturbation theory and construct the dilatation charge. Then, by using the Ward-Takahashi (WT) identity involved with the charge, we derive two expressions of the spectral indices for the scalar and the tensor perturbations without any ansatzes nor approximations. By construction, both expressions are \textit{exact} and  applicable to almost all wavelengths and all times for generic models. One of the two expressions is suitable for the slow roll expansion, while the other is suitable for numerical calculations.
Since we ignore the quantum effects, our results remain at the classical level.   
It is still enough, however, to compare with the previous results for linear perturbations.  
%ゆらぎがスクイーズされて古典化すると考えられていることや、
%
We note that the dilatation transformation and the charge here are entirely different from 
those which are used to relate the three-point correlators with the two-point correlators \cite{Weinberg:2003sw, Baumann:2011su, Assassi:2012zq, Goldberger:2013rsa, Hinterbichler:2013dpa, Tanaka:2017nff}. 
In the previous works, the authors performed large gauge transformations that are included into the  diffeomorphism, which were applied for the soft theorems. Here, we perform instead the dilatation transformation that is included into the conformal transformation regardless of the wavelengths of the perturbations.

%contents of this article
This article is organized as follows. In Sec. \ref{sec2}, firstly, we give generic actions for the linear perturbations. Next, we perform the dilatation transformation, and derive the current (non-)conservation law and the charge. Then, constructing the WT identity, we derive the exact expressions for the spectral indices. In Sec. \ref{sec3}, we expand the spectral indices by the slow roll parameters up to the second order. We derive rigorous and generic expressions 
for the spectral indices and their runnings, and compare the results with the previous ones. 
In Sec. \ref{sec4}, we analyze one of the exact expressions in detail and perform numerical calculations of the scalar spectral index for the Starobinsky model \cite{Starobinsky:1980te}. We compare the result with the slow roll expansion and the large $N_*$ expansion \cite{Roest:2013fha, Martin:2016iqo}. In the large $N_*$ expansion, we will expand the time-dependent parameters by the e-folding number at the horizon crossing time $N_*$. Sec. \ref{sec5} is devoted to the summary. We add two appendices. In Appendix \ref{A}, we deform one of the exact expressions for the indices to the other exact expression. In Appendix \ref{B}, we consider two specific models, power-law inflation \cite{Lucchin:1984yf} and ultra slow roll inflation \cite{Kinney:2005vj}, for the slow roll expansion.
We use the unit $8\pi G= c=\hbar = 1$ throughout this article.

%section2============================================================== 
\section{Ward-Takahashi identity}\label{sec2}

We consider single field inflation models in a generic way. Instead of specifying a model, we just give a generic action $S[\phi, g_{\mu\nu}]$, which consists of the scalar field $\phi(x)$ and the metric field $g_{\mu\nu}(x)$. We assume that the given theory possesses the 4-dimensional diffeomorphism invariance and has the flat Friedmann-Lema\^{\i}tre-Robertson-Walker background solution. For the single field inflation models,
the background dynamics is governed by the homogeneous part of the scalar field $\phi(t)$. To consider the cosmological perturbation theory around the background, we use the Arnowitt-Deser-Misner formalism and take the comoving gauge as follows:
%where the \RE{perturbation} of scalar field set to zero:
\begin{align}
\label{ }
&ds^2= -N^2dt^2 + {}^3g_{ij}(dx^i + N^idt)(dx^j+ N^jdt) \ ,    \nonumber  \\
    {}^3&g_{ij} = a^2{\rm e}^{2\zeta}[{\rm e}^\gamma]_{ij},\quad \partial_j\gamma_{ij} = 0 = \gamma_{ii}\ , \nonumber \\
 &\phi = \phi(t),\quad \delta\phi(x)= 0 \ .
\end{align}
$\zeta$ is the curvature perturbation and $\gamma_{ij}$ is the traceless-transverse modes of  tensor perturbation.
In this gauge, the generic action reduces to
\begin{equation}
\label{ }
S[\phi,\,g_{\mu\nu}] \rightarrow S[t,\, K_{ij},\, N,\, {}^3g^{ij},\,{}^3R_{ij},\,\cdots] \ ,
\end{equation}
where $K_{ij}$ and ${}^3R_{ij}$ are the extrinsic and intrinsic curvature, respectively, and the dots denote irrelevant higher derivative terms.
 
Following the derivation in \cite{Tanaka:2012wi, Saitou:2017xet}, %Solving all of constraints the theory has,
we can naturally obtain the generic free actions of scalar perturbation and tensor perturbation:
\begin{equation}
\label{ss}
S_s= \int dtd^3xa^3\left[ {\cal G}_s(t)\dot{\zeta}^2 - \frac{{\cal F}_s(t)}{a^2}(\nabla \zeta)^2 \right] \ ,
\end{equation}
\begin{equation}
\label{st}
S_t= \frac{1}{8}\int dtd^3xa^3\left[ {\cal G}_t(t)\dot{\gamma}_{ij}^2 - \frac{{\cal F}_t(t)}{a^2}(\nabla \gamma_{ij})^2 \right] \ ,
\end{equation}
where
%\begin{equation}
%\label{ }
%\gamma_{ij} = \sum v_{+,\times}{\rm e}_{ij},\quad {\rm e}_{ij}^2 = 1 \ ,
%\end{equation}
we restrict the perturbations to possess the Lorentz-invariant scaling. The dot denotes the partial derivative with respect to the cosmic time $t$. These generic free actions include all of Horndeski theory \cite{Kobayashi:2011nu, Horndeski:1974wa} and a part of Degenerate-Higher-Order-Scalar-Tensor theories \cite{Gleyzes:2013ooa, Gleyzes:2014dya, Gleyzes:2014qga, Gleyzes:2014rba, Langlois:2017mxy}.

We define the speeds of sound, the conformal times and the canonical perturbations for the scalar and the tensor perturbation respectively as
\begin{align}
\label{}
   &c_s^2 := \frac{{\cal F}_s}{{\cal G}_s}\ ,\nonumber \\
   \label{taus}
   &\tau_s := \int^t \frac{c_s}{a}dt\ ,  \\ 
    &u_s(x):= z_s\zeta(x),\quad z_s= \sqrt{2}a({\cal F}_s{\cal G}_s)^{1/4}\ , 
\end{align}
\begin{align}
\label{}
   &c_t^2 := \frac{{\cal F}_t}{{\cal G}_t}\ ,\nonumber \\
   \label{taut}
   &\tau_t := \int^t\frac{c_t}{a}dt\ ,  \\ 
    &u_t(x)\textbf{e}_{ij}:= z_t\gamma_{ij}(x), \quad z_t= \frac{a}{2}({\cal F}_t{\cal G}_t)^{1/4} \ .
\end{align}
%
%The definitions of conformal times include indefinite integrals.
Here, we defined the conformal times $\tau_{s,t}$ as the primitive functions of $c_{s,t}(t)/a(t)$, so that we set the constants of integration in Eqs.~(\ref{taus}) and (\ref{taut}) to be zero. In the region where the integrands $c_{s,t}(t)/a(t)$ are monotonic functions of the cosmic time $t$, there exist one-to-one correspondences between $\tau_{s,t}$ and $t$, and we can use the conformal times instead of the cosmic time.
$\textbf{e}_{ij}$ is a polarization tensor satisfied with $\textbf{e}_{ij}\textbf{e}^{ij}=2$. 
%The conformal times $\eta_{s,t}$ defined in the above ways still have ambiguities about the constants of integration. Then, we further introduce the following definition of times:
%\begin{align}
%\label{}
%    \eta_s&=: \tau_s - \tau_{s*}  ,\quad d\eta_s = d\tau_s \nonumber \\
%    \eta_t&=: \tau_t - \tau_{t*},\quad d\eta_t = d\tau_t \ .
%\end{align}
%The new conformal times $\tau_{s,t}$ are the generating functions of $c_{s,t}(t)/a(t)$ respectively, and $\tau_{s*,t*}$ are constants corresponding to the new conformal times at $t_{s*,t*}$
%We do not identify the constants of integration for the conformal times, and indeed they are irrelevant to our formulation and final results at all.  
Then, we rewrite the free action as
\begin{align}
\label{ }
    S_s&= \int d\tau_sd^3x \left[ \frac{1}{2}\left((\partial_{\tau_s}u_s)^2 - (\nabla u_s)^2\right) -\frac{1}{2}m_s^2u_s^2\right] \ , \\
    S_t&= \int d\tau_td^3x \left[ \frac{1}{2}\left((\partial_{\tau_t}u_t)^2 - (\nabla u_t)^2\right) -\frac{1}{2}m_t^2u_t^2\right]\textbf{e}^{ij}\textbf{e}_{ij}
\end{align}
where
\begin{align}
    m_{s}^2 &= -\frac{\partial_{\tau_s}^2z_s}{z_s}  \ , \nonumber \\
 \label{mass}
    m_{t}^2 &= -\frac{\partial_{\tau_t}^2z_t}{z_t} \ .
\end{align}
The free actions of canonical fields have the same forms, so that we can analyze their evolutions in the same manner. 

To derive the spectral indices in the Fourier space, first, we consider a dilatation transformation defined in the coordinate space as
\begin{align}
\label{dil}
&x_I^\mu \to \tilde{x}^{\mu}_I = {\rm e}^{-\alpha}x_I^\mu,\quad u_I(x_I)\to  \tilde{u}_I(\tilde{x}^\mu_I)={\rm e}^\alpha u_I(x_I) \ ,
\end{align}
where the label $I= \{s,\,t\}$, $x_I^\mu=(\tau_I,\,{\bf x})$ and $\alpha$ is a global constant parameter. 
Note that even if we redefine $\tau_I$ by shifting a finite constant from the present definitions (\ref{taus}) and (\ref{taut}), we can always obtain the same transformation as Eq.~(\ref{dil}) by acting a time translation to the redefined $\tau_I$. Thus, the definition of $\tau_I$ never affects the results below.
Under the dilatation transformation, the Lie derivatives of the field and the Lagrangian density become
\begin{align}
\label{}
    &\delta u_I(x_I)= (1+ x_I^\mu\partial_{I\mu})u_I(x_I) \ ,   \\
    &\delta {\cal L}_I= \partial_{I\rho}X_I^\rho + \theta_Iu_I^2  \ , \nonumber \\
     &X^\mu_I:= x_I^\mu{\cal L}_I \ , \nonumber \\
         \label{theta2} 
      &\theta_I:= m_I^2 +\frac{\tau_I}{2}\partial_{\tau_I} m_{I}^2 \ ,
\end{align}
where $\partial_{I\mu} := \partial/\partial x^\mu_I$. We do not take a contraction for the label $I$.
We define (non-)conserved current as
\begin{align}
\label{ }
j_I^\mu :&= \frac{\partial {\cal L}}{\partial(\partial_\mu u_I(x_I))}\delta u_I - X_I^\mu \nonumber \\
            &= x_I^\rho\,(\partial_I^\mu u_I\partial_{I\rho} u_I -\delta^\mu_\rho {\cal L}_I) + \frac{1}{2}\partial_I^{\mu}(u_I^2) \ .
\end{align}
By using the Euler-Lagrange equation for $u_I(x_I)$, we can easily find a current (non-)conservation law 
\begin{equation}
\label{cons}
\partial_{I\mu} j_I^\mu = \theta_Iu^2_I \ .
\end{equation}
$\theta_I$ represents the breaking size of dilatation invariance of the free action for $u_{I}(x_I)$. 
By integrating Eq.~(\ref{cons}) on the spacetime manifold, we can derive a relation between the dilatation charge $Q_I$ and the breaking term
\begin{align}
    Q_I(\tau_I)&:= \int d^3x j_I^0(\tau_I,\,{\bf x}) \ , \\
    \label{cd}
    Q_I(\tau_I)-Q_I(\tau_{I0})&= \int^{\tau_I}_{\tau_{I0}}d\tilde{\tau}_I\partial_{\tilde{\tau}_I}\int d^3x  j_I^0(\tilde{\tau}_I,\,{\bf x}) \nonumber \\
    &= \int^{\tau_I}_{\tau_{I0}}d\tilde{\tau}_I\int d^3x \tilde{\partial}_{I\mu}j_I^\mu(\tilde{\tau}_I,\,{\bf x}) \nonumber \\
    &= \int^{\tau_I}_{\tau_{I0}}d\tilde{\tau}_I\int d^3x\theta_Iu_I^2(\tilde{\tau}_I,\,{\bf x}) \ ,   
\end{align}
where the spatial integration covers all of the spatial hypersurface.
$\tau_{I0}$ is an arbitrary time so that the $\tau_{I0}$ dependence of the charge and of the last line of Eq.~(\ref{cd}) will cancel out each other. 
This implies that we can take $\tau_{I0}$ to $-\infty$ 
and extrapolate the behaviors of the perturbations and the background before the inflation to both sides of 
Eq.~(\ref{cd})
even if we do not know the actual behavior of the whole of system before the inflation. The contribution from the extrapolated part will cancel each other out on both sides.
We will extrapolate the field behaviors as in the Minkowski spacetime at the past infinity as usual.

Now, we canonically quantize the system. We define the conjugate momentum as
\begin{align}
\label{}
    \pi_I(x_I)= \frac{\partial {\cal L}_I}{\partial (\partial_{\tau_I}u_I(x_I))}
                  = \partial_{\tau_I}u_I(x_I) \ ,
\end{align}
and introduce the canonical commutation relations at the same time
\begin{align}
\label{}
   &[u_I(\tau_I, {\bf x}),\ \pi_I(\tau_I, {\bf y})]= i\delta^3({\bf x}-{\bf y}) \ , \\
    &[u_I(\tau_I, {\bf x}),\ u_I(\tau_I, {\bf y})]= [\pi_I(\tau_I, {\bf x}),\ \pi_I(\tau_I, {\bf y})]= 0 \ .
\end{align}
Then, we perform the Fourier expansion for the field
\begin{align}
\label{expand}
    u_I(x_I)= \int \frac{dk^3}{(2\pi)^{3/2}}\left[a_{I\bf k}u_{Ik}(\tau_I) + a_{I\bf k}^\dag u_{Ik}^*(\tau_I)\right] {\rm e}^{i{\bf k}\cdot({\bf x-y})}   \ ,
\end{align}
where $k=|\mathbf{k}|$.
The Heisenberg equation for $\pi_I$ require the mode function $u_{Ik}$ to be satisfied with the Mukhanov-Sasaki(MS) equation
\begin{align}
\label{MS}
\partial^2_{\tau_I}u_{Ik} + \left[k^2 + m_I^2(\tau_I)%- \frac{\partial_{\tau_I}^2z_I}{z_I}
\right]u_{Ik} = 0  \ .
\end{align}
By imposing a normalization for the Wronskian $W$,
\begin{equation}
\label{W}
W := (\partial_{\tau_I}u_{Ik})u^*_{Ik} - u_{Ik}(\partial_{\tau_I}u_{Ik}^*) = -i \ ,
\end{equation}
we can normalize the commutation relations between the creation and annihilation operators as
\begin{align}
\label{}
    &[a_{I\bf k},\, a^\dag_{J\bf q}]= \delta^3({\bf k}-{\bf q})\delta_{IJ} \ ,   \\
    &[a_{I\bf k},\, a_{J\bf q}]= [a^\dag_{I\bf k},\, a^\dag_{J\bf q}] =0  \ .
\end{align}
We define the free vacuum $|0\rangle$ as
\begin{align}
\label{}
    a_{I\bf k}|0\rangle= 0  \quad {\rm for}\ {}^\forall{\bf k} \ .
\end{align} 
In Eq. (\ref{MS}), if the time-dependent mass term dominates over the oscillation term, namely, $k^2\ll |m_I^2|$, 
the mode function has the general asymptotic solutions 
\begin{align}
\label{asy}
u_{Ik} &\rightarrow A_I(k)z_I + B_I(k)z_I\int^{\tau_I} \frac{d\tilde{\tau_I}}{z^2_I} \ .
\end{align}
$A_I$ and $B_I$ are the normalization constants dependent on the spatial momentum $k$ only. We may determine them by requiring the free vacuum corresponds to the Bunch-Davies vacuum. 
We call the former solution $z_I$ and the latter solution $z_I\int^{\tau_I} d\tilde{\tau}_I/z^2_I$ as the growing mode and the decaying mode respectively.
%When the growing mode grows faster than the decaying mode, we can omit the latter one.  

Then, 
using Eq.~(\ref{cd}), we write a WT identity for the two point function as
%\begin{align}
%\label{wt}
%    &\langle 0|[iQ_I(\tau_I), \ u_I(\tau_I,\,{\bf x})u_I(\tau_I,\, {\bf y})]|0\rangle \nonumber \\
%     =  &\langle 0|[i\int^{\tau_I} d\tilde{\tau}_I\int d^3z \theta_Iu_I^2 \, + b_I,\ u_I(\tau_I,\,{\bf x})u_I(\tau_I,\, {\bf y})]|0\rangle \nonumber \\ 
%     = &\langle 0|[i\int^{\tau_I} d\tilde{\tau}_I\int d^3z \theta_Iu_I^2,\ u_I(\tau_I,\,{\bf x})u_I(\tau_I,\, {\bf y})]|0\rangle
%\end{align}
\begin{align}
\label{wt}
    &\langle 0|[iQ_I(\tau_I), \ u_I(\tau_I,\,{\bf x})u_I(\tau_I,\, {\bf y})]|0\rangle \nonumber \\
     =  &\langle 0|[iQ_I(-\infty),\ u_I(\tau_I,\,{\bf x})u_I(\tau_I,\, {\bf y})]|0\rangle \nonumber \\ 
     + &\langle 0|[i\int^{\tau_I}_{-\infty} d\tilde{\tau}_I\int d^3z \theta_Iu_I^2,\ u_I(\tau_I,\,{\bf x})u_I(\tau_I,\, {\bf y})]|0\rangle \ ,
\end{align}
where we take $\tau_{I0}=-\infty$.
In general, of course, the above identity does not hold for an interacting field since the naive relation (\ref{cd}) will be changed by quantum corrections. However, at least for the free field, we can still use the classical one (\ref{cd}) as a relation between the quantum operators. 
We can calculate both sides of identity straightforwardly. The left hand side becomes
\begin{align}
\label{}
    &\langle 0|[iQ_I(\tau_I),\ u_I(\tau_I,\,{\bf x})u_I(\tau_I,\, {\bf y})]|0\rangle  \nonumber \\
    = &\int \frac{d^3k}{(2\pi)^3}\left[ -k\partial_k +\tau_I\partial_{\tau_I} -1\right]|u_{Ik}|^2{\rm e}^{i\mathbf{k}\cdot(\mathbf{x}-\mathbf{y})}  \nonumber \\ 
    &+\int\frac{dkd\Omega}{(2\pi)^3}\partial_k\left( k^3|u_{Ik}|^2{\rm e}^{i\mathbf{k}\cdot(\mathbf{x}-\mathbf{y})}\right) \ ,
\end{align}
where $d\Omega$ is a infinitesimal solid angle in the momentum space.
%We discarded the boundary terms of integration.
%\begin{equation}
%\label{ }
%Q_D = \int d^3x \left\{ \tau {\cal H} + x^i \partial_\tau u \partial_i x + u\partial_\tau u\right\}
%\end{equation}
%
On the other hand,
after several integration by parts, the first term of the right hand side becomes
\begin{align}
\label{}
    &\langle0|[iQ_I(\tau_{I0}=-\infty), \ u_I(\tau_I,\,{\bf x})u_I(\tau_I,\, {\bf y})]|0\rangle \nonumber \\
    &= \int d^3k \bigg\{i\tau_{I0}\bigg[(\partial_{\tau_I}u_{Ik0})^2u_k^{*2} 
        - (\partial_{\tau_I}u_{Ik0}^{*})^{2}u_k^2  \nonumber \\
     &\quad +(k^2+m_{I0}^{2})(u_{Ik0}^{2}u_k^{*2} - u_{Ik0}^{*2}u_k^2) \bigg] 
        \nonumber  \\
    &\quad -i(u_{Ik0}\partial_{\tau_I}u_{Ik0}u_k^{*2} - u_{Ik0}^{*}
        \partial_{\tau_I}u_{Ik0}^{*}u_k^2) \nonumber \\
    &\quad+ (\partial_{\tau_I}u_{Ik0}k\partial_ku_{Ik0}
         -u_{Ik0}k\partial_k\cdot\partial_{\tau_I}u_{Ik0})u_k^{*2}  \nonumber \\
    &\quad - (\partial_{\tau_I}u_{Ik0}^{*}k\partial_ku_{Ik0}^{*}
       -u_{Ik0}^{*}k\partial_k\cdot \partial_{\tau_I}u_{Ik0}^{*})u_k^2\bigg\} \nonumber \\
    &\quad \times {\rm e}^{i\bf{k}\cdot\bf(x-y)} \ ,
\end{align}
where the quantities with the subscript $0$ are evaluated at the time $\tau_{I0}=-\infty$.
%\RS{インフレーション前でも場がMSeqに従うとして（多分こんな感じに場の振る舞いを仮定しておかないと、ユニタリティーとかが崩れるかもしれんから、完全に任意には取れないと思う。少なくともロンスキアンの条件は満たさないと演算子展開が微妙になるし。）、マスも無限過去で０になるように宇宙の発展を仮想的に外挿入しておくと、以下のもんが得られる。}
%If we identify the vacuum $|0>$ to the Bunch-Davies vacuum, 
We extrapolate the mode function as it obeys Eq.~(\ref{MS}) even before the inflation. 
We also extrapolate 
the mass at the past infinity as it approaches to zero faster than $1/\tau_I$ based on the estimation $m_I^2\sim -1/\tau_I^2$. Then, 
we obtain
\begin{align}
\label{uasy}
    &\tau_{I0}m_{I0}^2|_{\tau_{I0}=-\infty}=0 \ , \nonumber \\ 
    &u_{Ik0}= \frac{1}{\sqrt{2k}}{\rm e}^{-ik\tau_{I}}\bigg|_{\tau_I = -\infty} \ .  
\end{align} 
We note that these extrapolations do not need to follow the actual history of the universe before the inflation since the $\tau_{I0}$ dependence will cancel each other on both sides of WT identity.
It is sufficient to fix the mass and the mode function at past infinity, with which one can perform the calculation straight-forwardly. 
%\RE{It is sufficient to set the quantities at the past infinity calculable}.
Then, we find the first term in the right hand side of Eq. (\ref{wt}) includes a surface term only:
\begin{align}
\label{ }
&\langle 0|[iQ_I(-\infty), \ u_I(\tau_I,\,{\bf x})u_I(\tau_I,\, {\bf y})]|0\rangle \nonumber \\
   = & i \int\frac{dkd\Omega}{(2\pi)^3}\partial_k\left[ ( k^3\partial_{\tau_I}u_{Ik0}\cdot u_{Ik0}u_{Ik}^{*2} - {\rm c.c.})\,{\rm e}^{i\mathbf{k}\cdot(\mathbf{x}-\mathbf{y})}\right] \ ,
\end{align}
where c.c. implies the complex conjugate.
%This implies that the tree-level free action have an asymptotic scale invariance at the past infinity since for the Bunch-Davies vacuum, the field behaves as a massless free scalar field on Minkowski spacetime, which possesses the scale invariance, at the past infinity. 
%
The breaking term becomes
\begin{align}
\label{}
    &\langle 0|[i\int^{\tau_I}_{-\infty} d\tilde{\tau}_I\int d^3z \theta_Iu_I^2,\ u_I(\tau_I,\,{\bf x})u_I(\tau_I,\, {\bf y})]|0\rangle   \nonumber \\
    = &2i\int \frac{d^3k}{(2\pi)^3} \int^{\tau_I}_{-\infty} d\tilde{\tau}_I\nonumber \\
  \times &\theta_I(\tilde{\tau}_I)\left[u_k^2(\tilde{\tau}_I)u_{Ik}^{*2}(\tau_I)-u_{Ik}^{*2}(\tilde{\tau}_I)u_{Ik}^2(\tau_I)\right]{\rm e}^{i\mathbf{k}\cdot(\mathbf{x}-\mathbf{y})} \ .
\end{align}
Then, we consider the inverse Fourier transformation for the WT identity (\ref{wt}).
Multiplying ${\rm e}^{-i(\mathbf{p}\cdot\mathbf{x}+\mathbf{q}\cdot\mathbf{y})}$ and taking  integrations for $\mathbf{x}-$ and $\mathbf{y}-$space, we obtain
\begin{align}
\label{Fcom}
     &\left[ -p\partial_p +\tau_I\partial_{\tau_I} -1\right]|u_{Ip}|^2\cdot (2\pi)^3\delta^3(\mathbf{p}+\mathbf{q}) \nonumber   \\
   &+\int dkd\Omega\, \partial_k\left[ (2\pi)^3k^3|u_{Ik}|^2\delta^3(\mathbf{k}-\mathbf{p})\delta^3(\mathbf{k}+\mathbf{q})\right] \nonumber  \\
   =&\int dkd\Omega\, \partial_k\left[ (2\pi)^3k^3 (\partial_{\tau_I}u_{Ik0}\cdot u_{Ik0}u_{Ik}^{*2} - {\rm c.c.})\right. \nonumber \\
   &\qquad\quad \left. \cdot \delta^3(\mathbf{k}-\mathbf{p})\delta^3(\mathbf{k}+\mathbf{q})\right] \nonumber  \\
   +&2i\int^{\tau_I}_{-\infty}d\tilde{\tau}_I\theta_I(\tilde{\tau}_I)\left[u_{Ip}^2(\tilde{\tau}_I)u_{Ip}^{*2}(\tau_I)-u_{Ip}^{*2}(\tilde{\tau}_I)u_{Ip}^2(\tau_I)\right] \nonumber \\
   &\cdot (2\pi)^3\delta^3(\mathbf{p}+\mathbf{q}) \ .
\end{align} 
%
%・座標空間でのward-idは定義可能とする。で、そのフーリエ変換も定義できるとする。
For the surface terms (the second and the third terms), we consider two edge modes, $k=0,\,\infty$. From the structure of the Dirac's delta function, we find that the surface terms contribute only when $p=0,\,\infty$ and $\mathbf{p}+\mathbf{q}=0$, while the remaining terms exist as long as $\mathbf{p}+\mathbf{q}=0$. Therefore, besides $p=0$ and $\infty$ modes, the following relation must be satisfied:
\begin{align}
\label{Wid}
&\left[ p\partial_p -\tau_I\partial_{\tau_I} +1\right]|u_{Ip}|^2 \nonumber \\
&= -2i\int^{\tau_I}_{-\infty}d\tilde{\tau}_I\theta_I(\tilde{\tau}_I)\left[u_{Ip}^2(\tilde{\tau}_I)u_{Ip}^{*2}(\tau_I)-u_{Ip}^{*2}(\tilde{\tau}_I)u_{Ip}^2(\tau_I)\right] \ .
\end{align} 
We regard this relation as the WT identity in the momentum space, which can be applicable for all modes except for $p=0$ and $\infty$ modes.
In the next section, we will see that the identity (\ref{Wid}) perfectly reproduces the previous results of the spectral index for the canonical scalar field \cite{Gong:2001he} without any ansatzes for the parameters. 
We comment also that a massless free scalar field on Minkowski spacetime or de Sitter spacetime is formally satisfied with the WT identity (\ref{Wid}) even for $p=0$ and $\infty$ modes.
%the WT identity (\ref{Wid}) is formally satisfied holds for a massless free scalar field on Minkowski spacetime or de Sitter spacetime .
To specify a rigorous treatment for the surface terms lies beyond the scope of this paper, but it might be connected with some symmetries of vacuum like the residual gauge symmetries \cite{Weinberg:2003sw, Baumann:2011su, Assassi:2012zq, Goldberger:2013rsa, Hinterbichler:2013dpa, Tanaka:2017nff}. We put this issue for future investigation.

We define the tree level power spectra of the scalar perturbation and the tensor perturbation as 
\begin{align}
\label{ps}
P_s &:= \frac{k^3}{2\pi^2}\left|\frac{u_{sk}}{z_s}\right|^2 \ , \nonumber \\
P_t&:= 2\frac{k^3}{2\pi^2}\left|\frac{u_{tk}}{z_t}\right|^2 \ . 
\end{align}
The scalar and tensor spectral indices are defined as
\begin{align}
\label{nso}
n_s-1 &:= \frac{d{\rm ln}P_s}{d{\rm ln}k} = \frac{k\partial_k P_s}{P_s} \ , \\
n_t&:= \frac{d{\rm ln}P_t}{d{\rm ln}k} = \frac{k\partial_k P_t}{P_t}\ .
\end{align}
These relations imply
\begin{align}
\label{ku}
k\partial_k|u_{sk}|^2 &= (n_s-4)|u_{sk}|^2 \ , \\
\label{kt}
k\partial_k|u_{tk}|^2 &= (n_t-3)|u_{tk}|^2 \ .
\end{align}
To derive the explicit expressions of the spectral indices, we substitute Eqs.~(\ref{ku}) and (\ref{kt}) to the WT identity~(\ref{Wid}). Then, we finally obtain
\begin{align}
\label{ns}
  n_I&= {\cal N}_I+ 2+2\tau_I\frac{\partial_{\tau_I}|u_{Ik}|}{|u_{Ik}|} + \Theta_I \\
%  n_s&= 3+2\tau_s\frac{\partial_{\tau_s}|u_{sk}|}{|u_{sk}|} + \Theta_s \ , \\
%           \label{nt}
%  n_t &= 2+2\tau_t\frac{\partial_{\tau_t}|u_{tk}|}{|u_{tk}|} +\Theta_t \ , \\
         \label{Theta}
\Theta_I&:= 4{\rm Im}\left[\frac{u_{Ik}^{*2}(\tau_I)}{|u_{Ik}(\tau_I)|^2}\int^{\tau_I} _{-\infty}
\theta_I(\tilde{\tau}_I)u_{Ik}^2(\tilde{\tau}_I)d\tilde{\tau}_I\right] \ .
\end{align}
where ${\cal N}_s= 1$, ${\cal N}_t= 0$.
We can deform $n_I$ into another form which is useful for a numerical calculation. 
Performing several integrations by parts with use of Eqs.~(\ref{MS}) and (\ref{W}), we obtain
\begin{align}
   \label{npsi}
   &n_I = {\cal N}_I +1 - 4{\rm Im}\left[\frac{\psi_{Ik}^{*2}}{|\psi_{Ik}|^2}\int^{\tau_I}_{-\infty(1+i\varepsilon)}
    z_I^2(\partial_{\tilde{\tau}_I}\psi_{Ik})^2d\tilde{\tau}_I\right] \ , \\
%    &n_s = 2 - 4{\rm Im}\left[\frac{\psi_{sk}^{*2}}{|\psi_{sk}|^2}\int^{\tau_s} z_s^2(\partial_{\tau_s}\psi_{sk})^2d\tilde{\tau}_s\right] \ , \\
%    \label{nsn}
%    &n_t = 1 - 4{\rm Im}\left[\frac{\psi_{tk}^{*2}}{|\psi_{tk}|^2}\int^{\tau_t} z_t^2(\partial_{\tau_t}\psi_{tk})^2d\tilde{\tau}_t\right] \ , \\
\label{psik}
    &\psi_{Ik}:= \frac{u_{Ik}}{z_I} \ ,
\end{align}
where we take the contour the same as the $i\varepsilon$ prescription.
See Appendix A for the technical detail of this deformation.

Here, we stress that all of the above results (\ref{ns})-(\ref{npsi}) are \textit{exact} expressions of the tree level spectral indices without any ansatzs nor approximations.  Basically, we can use them for all time and all wavelengthes except for $k=0$ and $\infty$ modes, not restricted to the super-horizon scales. 
%Indeed, at $\tau_I = -\infty$, we obtain
%\begin{equation}
%\label{ }
%n_s-1 = 2+ 0 + 0= 2,\quad n_t = 2+0+0 = 2\ ,
%\end{equation}
%which are consistent.
We can apply Eqs.~(\ref{ns})-(\ref{npsi}) to \textit{all} of the single field inflation models which possess the Lorentz-invariant scaling even if the background and the perturbations evolve in highly nontrivial ways.

%section3==========================================================
\section{Slow roll expansion}\label{sec3}

In this section, we expand the exact expression (\ref{ns}) by the slow roll parameters
up to the second order and compare the results with the previous works. First, we introduce the e-folding number $n$ as
\begin{align}
\label{n}
    n&:= {\rm ln}\left(\frac{a}{a_{\rm in}}\right)\ ,   \\
    a&= a_{\rm in}{\rm e}^n \ ,  \nonumber
\end{align} 
where $a_{\rm in}$ is an initial finite value of the scale factor at a certain time, and thus $n_{\rm in} = 0$. If the universe always expands, we can use $n$ as a clock. Next, we define the generic slow roll parameters following \cite{Kobayashi:2011nu} as
\begin{align}
\label{srpara}
    \epsilon_1&:= -\frac{d\,{\rm ln}H}{dn},\quad \epsilon_{i+1}:= \frac{d\,{\rm ln\epsilon_i}}{dn} \ ,  \nonumber\\
    f_{I1}&:= \frac{d\,{\rm ln}{\cal F}_I}{dn}, \quad f_{Ii+1}:= \frac{d\,{\rm ln}f_{Ii}}{dn} \ , \nonumber\\
    g_{I1}&:= \frac{d\,{\rm ln}{\cal G}_I}{dn}, \quad g_{Ii+1}:= \frac{d\,{\rm ln}g_{Ii}}{dn}   \ ,
\end{align}
where $H := d\,{\rm ln}a/dt$ is the Hubble parameter and ${\cal F}_I$ and ${\cal G}_I$ are the coefficient functions appearing in the original free actions (\ref{ss}) and (\ref{st}).
%
%As the simplest example, we calculate $n_I$ and $\alpha_I$ for the case $z_s = M_{\rm pl}a\sqrt{2\epsilon_1}$, $z_t= M_{\rm pl}a/2$ and $c_s=c_t=1$.
%ここでコンセントレイトしないで、一般のzについての定式を書く。
%SRパラの定義も書く。
%e-foldも導入
By using $n$ and the slow roll parameters, we can expand the conformal time as    
%\begin{equation}
%\label{tau}
%\tau_I =  -\frac{1+\epsilon_1\epsilon_2+ O(\epsilon_i^3)}{aH(1-\epsilon_1)} \ .
%\end{equation} 
\begin{align}
\label{}
    \tau_I&= \int^n \frac{c_I}{aH}d\tilde{n} \nonumber    \\
             &= -\frac{c_I}{aH}\sum_{i=0}^\infty \delta_{Ii} \ ,\nonumber  \\
    \delta_{I0}&= 1, \nonumber \\
    \delta_{I1}&= \epsilon_1+\frac{1}{2}(f_{I1}-g_{I1}) \ , \nonumber \\
    \delta_{Ii} &:= \delta_{I1}\delta_{Ii-1} + \delta_{Ii-1,n} \quad {\rm for}\ i\geq 2\ . 
\end{align}
where `$,n$' denotes the differentiation with respect to $n$.
The subscript $i$ on $\delta_{Ii}$ corresponds to the order of slow roll parameters.
Using them, up to the second order, the breaking size $\theta_I$ becomes
%\begin{equation}
%\label{ }
%\theta_{s}\simeq \frac{3\left(\epsilon_1\epsilon_2+\frac{1}{2}\epsilon_2\epsilon_3\right)}{2\tau_s^2},\quad \theta_{t}\simeq \frac{3\epsilon_1\epsilon_2}{2\tau_t^2} \ .
%\end{equation}
\begin{align}
\label{t2}
    \theta_I = \frac{3}{2\tau_I^2}\left[ \epsilon_1\epsilon_2 + \frac{1}{4}
                    (3f_{I1}f_{I2}- g_{I1}g_{I2}) + O(\epsilon_i^3)\right] \ .
\end{align}
We denoted higher order terms of the slow roll parameters  than the second order by $O(\epsilon_i^3)$. 
%because it represents the breaking size of the scale invariance of the free action.
To evaluate $\Theta_I$ (\ref{Theta}) up to the second order of slow roll parameters, 
we expand the mode function also as follows:
\begin{align}
\label{}
    u_{Ik}(\tau_I)= \sum_{i=0}^\infty u_{Ik}^{(i)}(\tau_I) \ , 
\end{align}
where $u_{Ik}^{(i)}$ is the $i$-th order quantity of slow roll parameters.
Since $\theta_{I}$ is already of the second order, we need the zeroth order term of the mode function $u_{Ik}^{(0)}$ only. In the slow roll expansion, the zeroth order term of mode function in the quasi-de Sitter spacetime must be given by that in the exact de Sitter spacetime, so we can replace the mode function in $\Theta_I$ with
\begin{equation}
\label{u0}
u_{Ik} \approx u_{Ik}^{(0)} = \frac{-i}{\sqrt{2k^3}\tau_I}(1+ ik\tau_I){\rm e}^{-ik\tau_I}  \ .
\end{equation}
Substituting Eqs.~(\ref{t2}) and (\ref{u0}) into Eq.~(\ref{Theta}), we obtain
%\begin{align}
%\label{ }
%\Theta_s &\simeq \left\{ {\rm Re}[-8{\rm Ei}(-2ik\tau_s)-6\Gamma(0,2ik\tau_s)]+4\right\} \left(\epsilon_1\epsilon_2+\frac{1}{2}\epsilon_2\epsilon_3\right) \nonumber \\
%&\simeq -2(C +{\rm ln}(-k\tau_s))\left(\epsilon_1\epsilon_2+\frac{1}{2}\epsilon_2\epsilon_3\right) \ ,\nonumber \\
%\Theta_t &\simeq \left\{ {\rm Re}[-8{\rm Ei}(-2ik\tau_t)-6\Gamma(0,2ik\tau_t)]+4\right\} \epsilon_1\epsilon_2 \nonumber \\
%&\simeq -2(C +{\rm ln}(-k\tau_t))\epsilon_1\epsilon_2 \ , \nonumber \\
%C&:= \gamma_E +{\rm ln}2-2 \ ,
%\end{align}
\begin{align}
\label{tsr}
    \Theta_I&=  -2\left[ \epsilon_1\epsilon_2 + \frac{1}{4}
                    (3f_{I1}f_{I2}- g_{I1}g_{I2})\right] \nonumber \\
                    &\ \ \quad \times\left[C + {\rm ln}(-k\tau_I) + O(k\tau_I) \right] + O(\epsilon_i^3)  \ ,    \\
    C:&= \gamma_E +{\rm ln}2-2 \ , \nonumber 
\end{align}
where $\gamma_E$ is the Euler-Mascheroni constant and $C\approx-0.729637$. 
In the calculation, we treated the second order product of slow roll parameters as a constant since its variation becomes of the third order.  

Now, we focus on the case where the asymptotic solutions of mode function (\ref{asy}) are governed by the growing mode\footnote{For other specific cases, see Appendix B.}  
\begin{equation}
\label{uas}
u_{Ik}\rightarrow A_I(k)z_I \quad {\rm when} \ \ |k\tau_I|\rightarrow 0 \ . 
\end{equation}
In this case, $\psi_{Ik}= u_{Ik}/z_I$ and the spectra (\ref{ps}) depend on the spatial momentum only, 
\begin{align}
\label{}
    \psi_{Ik}&\rightarrow A_I(k) \ ,\nonumber  \\
    P_s&\rightarrow \frac{k^3}{2\pi^2}|A_s(k)|^2,\quad P_t\rightarrow \frac{k^3}{\pi^2}|A_t(k)|^2,  
\end{align}
which implies that they almost conserve on the super-horizon scales.
The asymptotic values of spectral indices become 
%where we omit $O(k\tau_I)$ term in $\Theta_I$ since it decreases exponentially to an extremely small value.
\begin{align}
\label{nsr}
    &n_I|_{|k\tau_I|\rightarrow 0} \nonumber \\
    &= {\cal N}_I -2\epsilon_1 -\frac{1}{2}(3f_{I1}-g_{I1}) \nonumber  \\
    &-\left\{ 2\epsilon_1^2 + \frac{\epsilon_1}{2}(5f_{I1}-3g_{I1}) +\frac{1}{4}(f_{I1}-g_{I1})
        (3f_{I1}-g_{I1}) \right. \nonumber \\
     &+ (2C+2)\epsilon_1\epsilon_2 +\frac{3C+2}{2}f_{I1}f_{I2} -\frac{C+2}{2}g_{I1}g_{I2} 
        \nonumber \\
     &\left. + \left[2\epsilon_1\epsilon_2 + \frac{1}{2}(3f_{I1}f_{I2}-g_{I1}g_{I2})\right] {\rm ln}(-k\tau_I) \right\} \nonumber \\
     &+ O(\epsilon_i^3)  \ .
\end{align}
%
%where ${\cal N}_s= 1,\ {\cal N}_t = 0$.   
We note that although Eq.~(\ref{nsr}) seems to depend on the time, actually it \textit{does not} since the asymptotic values of spectra depend on the spatial momentum only.
Therefore, we can change the time in Eq.~(\ref{nsr}) to an arbitrary time. If we change it to the horizon crossing time for the mode $k_*$, 
\begin{equation}
|\tau_I| \Rightarrow \frac{1}{a_*H_*} = \frac{1}{k_*}\ ,
\end{equation}      
we finally obtain the asymptotic values of spectral indices in a generic form as
\begin{align}
\label{nstar}
    &n_I(k)|_{|k\tau_I|\rightarrow 0} \nonumber \\
    & = {\cal N}_I -2\epsilon_{1*} -\frac{1}{2}(3f_{I1*}-g_{I1*}) \nonumber  \\
    &-\left\{ 2\epsilon_{1*}^2 + \frac{\epsilon_{1*}}{2}(5f_{I1*}-3g_{I1*}) \right. \nonumber \\
    &+\frac{1}{4}(3f_{I1*}^2-4f_{I1*}g_{I1*}+ g_{I1*}^2) \nonumber \\
        %(f_{I1*}-g_{I1*})(3f_{I1*}-g_{I1*}) \right. \nonumber \\
     &+ (2C+2)\epsilon_{1*}\epsilon_{2*} +\frac{3C+2}{2}f_{I1*}f_{I2*} -\frac{C+2}{2}g_{I1*}g_{I2*} 
        \nonumber \\
     &\left. + \left[2\epsilon_{1*}\epsilon_{2*} + \frac{1}{2}(3f_{I1*}f_{I2*}-g_{I1*}g_{I2*})\right] {\rm ln}\left(\frac{k}{k_*}\right) \right\} \nonumber \\
     &+ O(\epsilon_{i*}^3)  \ .
\end{align}
All of the slow-roll parameters are evaluated at the horizon crossing time for the mode $k_*$ when $ k_*= a_*H_*(n_*)$. The logarithmic term represents the deviation between the modes $k$ and $k_*$, and it disappears if we choose $k=k_*$. 
We can safely ignore the higher order terms denoted by $O(\epsilon_{i*}^3)$ for the models that all of the slow-roll parameters are small enough at the horizon crossing time. We stress that we evaluate the asymptotic values of spectral indices, say, the values at the end of inflation, by using the slow roll parameters evaluated at the horizon crossing time. This is the ``horizon crossing formalism" itself.
 
To derive the runnings of spectral indices defined as
\begin{equation}
\alpha_I := \frac{dn_I}{d{\rm ln}k}\ ,
\end{equation}
 we choose $k= k_*$, so that we can relate the spatial momentum to the horizon crossing time through $k=k_*=a_*H_*(n_*)$. By this, we get a relation between differentiations
\begin{align}
\label{}
    \frac{d}{d{\rm ln}k}&= \frac{d}{d{\rm ln}(a_*H_*)} 
    = \frac{1}{1-\epsilon_{1*}}\frac{d}{dn_*}   \ .
\end{align}
%
%Differentiating Eq.(\ref{nstar}) with respect to ${\rm ln}k_*$, we can obtain a generic expression for the running of spectral indices. 
%%
%If the field evolves as $u_{Ik}\rightarrow A_Iz_I$ at $\tau_I \rightarrow -0$, neither the power spectrum nor $n_I$ depend on the time at $\tau_I \rightarrow -0$. In such a case, we obtain 
%\begin{equation}
%\label{as}
%\alpha_I|_{-k\tau_I\rightarrow 0} = k\partial_k\Theta_I|_{-k\tau_I\rightarrow 0} \ .
%\end{equation}
%%
%If we expand $\Theta_I$ up to the second order of slow-roll parameters, 
%%for $z_t= M_{pl}a/2$ and $c_t=1$
%from Eq. (\ref{ns2}), we obtain
Then, we can obtain the asymptotic values of the runnings of spectral indices up to the third order
\begin{align}
\label{astar}
&\alpha_I(k_*)|_{|k\tau_I|\rightarrow 0} \nonumber \\
     &=-2\epsilon_{1*}\epsilon_{2*} - \frac{1}{2}(3f_{I1*}f_{I2*}-g_{I1*}g_{I2*}) \nonumber \\
     &-6\epsilon_{1*}^2\epsilon_{2*} -2\epsilon_{1*}(2f_{I1*}f_{I2*}-g_{I1*}g_{I2*}) \nonumber \\
     &-\frac{\epsilon_{1*}\epsilon_{2*}}{2}(5f_{I1*}-3g_{I1*}) -(2C+2)(\epsilon_{1*}\epsilon_{2*}^2
         +\epsilon_{1*}\epsilon_{2*}\epsilon_{3*}) \nonumber \\
     &-\frac{3}{2}f_{I1*}^2f_{I2*} + f_{I1*}f_{I2*}g_{I1*} + f_{I1*}g_{I1*}g_{I2*} -\frac{1}{2}g_{I1*}^2g_{I2*} \nonumber \\
     &-\frac{3C+2}{2}(f_{I1*}f_{I2*}^2+f_{I1*}f_{I2*}f_{I3*}) \nonumber \\
     &+ \frac{C+2}{2}(g_{I1*}g_{I2*}^2+g_{I1*}g_{I2*}g_{I3*}) \nonumber \\
     &+ O(\epsilon_{i*}^4) \ ,
\end{align}
where all of the slow roll parameters are evaluated at the horizon crossing time for the mode $k=k_*$.

As the simplest example, we consider the following case
%\begin{align}
%\label{}
%    S&= \int d^4x \left[ \frac{1}{2}R - \frac{1}{2}g^{\mu\nu}\partial_\mu\phi\partial_\nu\phi -V(\phi)\right] \ ,
%\end{align}
%where $R$ is the Ricci scalar and $V$ is the potential. For this model, the coefficients of scalar perturbation are given by
\begin{align}
\label{}
    {\cal F}_s&= {\cal G}_s = \epsilon_1  \ , \nonumber \\
    {\cal F}_t&= {\cal G}_t = 1  \ ,
\end{align}
which is obtained for the canonical scalar field models with the potential.  Using Eq.~(\ref{nstar}) and (\ref{astar}) for this case, up to the second order, we obtain  
\begin{align}
\label{nsv}
   n_s(k_*)|_{|k\tau_s|\rightarrow 0}
    &= 1 -2\epsilon_{1*} -\epsilon_{2*} -2\epsilon_{1*}^2 \nonumber   \\
    &\quad-(2C+3)\epsilon_{1*}\epsilon_{2*} - C\epsilon_{2*}\epsilon_{3*} \ , \nonumber \\
    n_t(k_*)|_{|k\tau_s|\rightarrow 0}
    &= -2\epsilon_{1*} -2\epsilon_{1*}^2 -2(C+1)\epsilon_{1*}\epsilon_{2*}  \ , \nonumber  \\ 
    \alpha_s(k_*)|_{|k\tau_s|\rightarrow 0}&= -2\epsilon_{1*}\epsilon_{2_*} 
                  -\epsilon_{2*}\epsilon_{3_*} \ , \nonumber \\
    \alpha_t(k_*)|_{|k\tau_s|\rightarrow 0}&= -2\epsilon_{1*}\epsilon_{2_*}  \ .            
\end{align}
These correspond to the previous results \cite{Gong:2001he, Casadio:2006wb}.   
In our formalism, however, the results are more rigorous and generic.
While the previous results rely on an ansatz for $z_I$ or the Wentzel-Kramers-Brillouin approximation for the mode function, our slow roll expansion does never need any ansatzes nor approximations.
 Also, we can easily derive the spectral indices and their runnings for the generic models without deriving the power spectra themselves. 
Further, in \cite{Casadio:2006wb}, they perform an uncertain Taylor expansion for the asymptotic values of spectral indices to evaluate them by the slow roll parameters at the horizon crossing time. We showed explicitly, however, that we do not need such a Taylor expansion and can evaluate the spectral indices just by applying the horizon crossing formalism. We derived the runnings of spectral indices up to the third order also.

%コメントする。これらの結果は、インフレーションエンドでSR近似が壊れることの影響をほぼ受けない。
%なぜなら、摂動がs-hで保存している場合、エンドの時刻でもずっとs-hスケールにいるはずであり、解はほとんど漸近解と変わらないからである。また、エンドの時刻は正確にはktau=0ではないから漸近解とは誤差があるが、エンド時にktauの値はおよそe-50~-60であり、ThetaSRの値から、その誤差もほとんど効かない。なので、実質的に、ここで求めた値をエンドの時の値として用いて良い。
We comment on tiny effects coming from around the end of inflation. In typical scenarios, the inflation ends when the slow roll parameters rapidly grow and reach to $O(1)$ values. Thus, around the end, the slow roll expansion would break down. However, this hardly affects the final results since in the case the perturbations freeze out on super-horizon scales, the mode function remains almost the same as the asymptotic value (\ref{uas}). 
On the other hand, the exact ending time of the inflation differs from $|k\tau_I|=0$, and thus the asymptotic values of spectral indices have errors, which are represented by the $O(k\tau_I)$ term in 
Eq.~(\ref{tsr}). Those errors are, however, extremely small since at the end of inflation, $|k\tau_I|\sim {\rm e}^{-50}$ for the observational window of the mode $k$. Therefore, in substance, we can use the asymptotic values of the spectral indices (\ref{nstar}) and their runnings (\ref{astar}) all the way to the end of the inflation.%as those of the inflation end.

We cannot apply, however, the slow roll expansion to the models in which the slow roll parameters grow above a certain limit nearby the horizon crossing time firstly, and then turn to small values again on the super-horizon scales. In these models, the mode function around the horizon crossing time behaves in a highly non-trivial way, and the contributions from $\Theta_I$ and/or $\tau_I$ deviate from the simple slow roll expansion significantly. We can further appreciate this by looking at the other expressions for the index (\ref{npsi}). We study their behaviors in the next section. 
%the perturbation stays near the horizon sizes at the first break of slow roll expansion and take a significant modification from the simple expansion.         
%しかし、ホライズンクロス近傍でスローロール近似が一定程度以上崩れたのち、再び良い近似となるような模型に対しては、上のSR近似の結果は正しくない。なぜなら、この時はモード関数のクロス付近での振る舞いが非自明になることで、τやΘからの寄与が単純なSR展開の値からずれてしまうからだ。これらのことはもう一つの表式に直すとよく理解できる。それは数値解のとこでやる。

%section4===============================================================
\section{Numerical analysis}\label{sec4}

In this section, we perform a numerical analysis for the Starobinsky model \cite{Starobinsky:1980te}. First, we give a numerical setup for the background and the perturbation of the model. We focus on the scalar perturbation only since at the present, the tensor spectral index is not yet observed.
Next, we compare the numerical result of slow roll expansion with the large $N_*$ expansion for the model. Then, 
we analyze the behavior of integration in Eq.~(\ref{npsi}), quoting the perturbation on the exact de Sitter background as a schematic example. We compare the numerical result of Eq.~(\ref{npsi}) for the scalar perturbation with the results of slow roll expansion and the large $N_*$ expansion also.

\subsection{Equations}
%In the following, we will assume the canonical scalar field models, i.e.  the models with the kinetic term given by $- \partial ^\mu  \phi \partial _{\mu} \phi /2$. 
We consider the Starobinsky model which has the following potential
\begin{equation}
V(\phi) = \frac{3}{4}M^4 \left ( 1- \mathrm{e}^{-\sqrt{\frac{2}{3}}\frac{\phi}{M_\mathrm{pl}}} \right )^2
\end{equation}
in the Einstein frame. $M$ is a model parameter which has the mass dimension $1$.
The Friedmann equations and the equation of motion of the scalar field are expressed as 
\begin{align}
3 H^2 &= \frac{1}{2} \dot \phi ^2 +V(\phi), \nonumber \\
\dot H &= - \frac{\dot \phi}{2}, \nonumber \\
\ddot \phi &+ 3H \dot \phi + V_{, \phi} =0 \ . \label{bges}
\end{align}
%
%where the dot denotes the derivative with respect to the cosmic time $t$.
If we use the e-folding number (\ref{n}) as a clock, 
we can rewrite Eqs.~(\ref{bges}) as 
\begin{align}
H^2 &= \frac{V}{3- \phi ^2_{, n}/2}, \label{feqn1} \\
\frac{H_{,n}}{H} &= - \frac{1}{2} \phi_{, n}^2, \label{feqn2} \\
\phi _{,nn} &+ \left ( 3+ \frac{H_{,n}}{H} \right ) \phi _{,n} + \frac{V_{, \phi}}{H^2} =0. \label{seqn}
\end{align}
%where we have set $M_\mathrm{pl} = 1$. 
Combining Eqs.~(\ref{feqn1})-(\ref{seqn}) gives the following single variable equation: 
\begin{equation}
\phi _{,nn} + \left ( \phi _{,n} + \frac{V_{, \phi}}{V} \right ) \left ( 3- \frac{1}{2} \phi _{,n} ^2 \right ) =0. \label{feq}
\end{equation}
We will use Eq.~(\ref{feq}) as a background equation. 
The behavior of the Hubble parameter will be obtained by substituting the solution of Eq.~(\ref{feq}) into Eq.~(\ref{feqn1}). 
The relevant slow roll parameters for this model are given as 
\begin{align}
\epsilon _1 &= \frac{1}{2} \phi _{,n} ^2, \label{e1}\\
\epsilon _2 &= -2 \left(3- \frac{1}{2} \phi _{,n} ^2%\epsilon _1
   \right) \left ( 1+ \frac{V_{, \phi}}{\phi_{,n}V} \right ) , \label{e2}\\
\epsilon _3 &= \phi _{, n} ^2 + \frac{\phi _{,n}}{2} \frac{V_{, \phi}}{V} + \frac{3V_{, \phi}}{\phi _{,n} V} 
+ \frac{\frac{V_{, \phi \phi}}{V}-\frac{V_{, \phi \phi}^2}{V^2}}{1+\frac{V_{, \phi}}{\phi _{,n} V}}.
\end{align}
While, the dynamics of the scalar perturbation $\psi_{sk}= u_{sk}/z_s$ is governed by the following equation: 
\begin{equation}
\psi _{sk,nn} + (3+ \epsilon _2 - \epsilon _1) \psi _{sk,n} + \frac{k^2}{a^2 H^2}\psi _{sk} =0\ , \label{lpe}
\end{equation}
which can be reduced from Eq.~(\ref{MS}). Here, $z_s$ for the model was given by $z_s = a \sqrt{2 \epsilon _1}$.% and $z_t = a/2$.

Then, we define the end of inflation as the time when
\begin{equation}
\label{ }
\epsilon_1(n_{\rm end}) = 1\ ,
\end{equation}
and we consider the scalar perturbation $\psi_{sk}$ which leave the horizon 54 e-folds before the end of inflation.
Namely, we determine the wavenumber $k$ of perturbation as  
$k_*=a_*H_*(n_*)%=a_\mathrm{in}\mathrm{e}^{n_*}H_*
$, where 
$n_*$ is satisfied with
\begin{equation}
N_* := \ln \left ( \frac{a_\mathrm{end}}{a_*} \right ) = n_\mathrm{end}-n_* =54\ . 
\end{equation}
%Here, subscript ``end'' expresses the e-folds of the end of inflation, defined as $\epsilon _1 (t_\mathrm{end}) =1$. 

\paragraph*{Initial Conditions}
We give the initial conditions and the mass parameter $M$ for the background as
\begin{align}
\phi _\mathrm{in} &= 6.246 M_\mathrm{pl}\ , \nonumber \\
\phi _{,n \mathrm{in}} &= -10^{-2} M_\mathrm{pl}, \label{ini2}\nonumber \\
M &= 1.419 \times 10^{-4} M_\mathrm{pl} \ ,
\end{align}
so that we get
%From Eqs.~(\ref{feqn1}), (\ref{e1}), (\ref{e2}), and (\ref{ini1})--(\ref{ini3}), we can derive 
\begin{align}
\label{}
    H_\mathrm{in}  &\simeq 10^{-8} M_\mathrm{pl} \sim \frac{E_\mathrm{GUT}^2}{M_\mathrm{pl}}\ , \nonumber \\
    \epsilon _{1 \mathrm{in}}&= 5.0 \times 10^{-5} \ ,  \nonumber \\
    \epsilon _{2 \mathrm{in}} &\simeq 10^{-2}  \ . 
\end{align}
%$\epsilon _{2 \mathrm{in}} = 10^{-2} H_\mathrm{in}  = 10^{-8} M_\mathrm{pl} \sim \frac{E_\mathrm{GUT}^2}{M_\mathrm{pl}}
%$, $M \simeq 1.419 \times 10^{-4} M_\mathrm{pl}$, 
%and $\epsilon _{1 \mathrm{in}}= 5.0 \times 10^{-5}$. 
The above initial conditions yield large e-folding number of the slow roll regime; %$\ln (a_\mathrm{end}/a_\mathrm{in}) \simeq 119.897$, that is, 
$n_{\rm end}\simeq 119.897$.  
The background approaches rapidly to the slow roll attractor regime where $\dot{\phi}\simeq -V_{,\phi}/3H$, so that after a few e-folds, the dynamics of $\phi(t)$ is hardly affected by the choice of initial values.

For the scalar perturbation, because of the problem of calculation costs, we will take $n_\mathrm{pin}$ as an initial e-folding number which satisfies $n_\mathrm{end}-n_\mathrm{pin}=64$. Then, we will redefine $n$ as $n- n_\mathrm{pin}$, so that in the new definition, we obtain $n_\mathrm{pin}=0$,
$n_*=10$ and $n_{\rm in}\simeq -55.897$. At the time $n_{\rm pin}=0$, the perturbation which leave the horizon at $n_*=10$ 
stays deep inside the horizon yet, so we approximate its initial values by those on the exact de Sitter background. Using Eq.~(\ref{u0}) with $\tau_s\simeq -1/aH$, we give the initial values of scalar perturbation at $n_{\rm pin}$ as
\begin{align}
\label{pini}
\psi _{sk}(n_{\rm pin}) & \simeq \frac{u_{sk}^{(0)}}{z_s} \bigg | _\mathrm{pin} \nonumber \\
&= \left [ \frac{1}{z_s \sqrt{2 k_*}}+i \frac{aH}{z_s \sqrt{2k_* ^3}} \right ] _\mathrm{pin} 
     \nonumber   \\
&=: [\pi_{sk} + i\sigma_{sk}]_{\rm pin}\ , \\
 \psi _{sk,n}(n_ \mathrm{pin}) 
%&= \psi _{Ik} \left ( -\frac{z_{I,n}}{z_I} + \frac{u_{Ik,n}}{u_{Ik}} \right ) \bigg | _\mathrm{in} \nonumber \\
& \simeq \left[ \pi _{sk} \left ( -\frac{z_{s,n}}{z_s}  + \frac{1}{1+k_* ^2/(a^2 H^2)}\right )\right.  \nonumber \\
&+ \left. \sigma _{sk} \frac{k_*^3/(a^3H^3)}{1+k_* ^2/(a^2 H^2)} \right]\bigg |_\mathrm{pin} 
    \nonumber \\
    &+ i \bigg \{ - \pi _{sk} \frac{k_*^3/(a^3H^3)}{1+k_* ^2/(a^2 H^2)} 
\nonumber \\
&+ \sigma _{sk} \left [ - \frac{z_{s,n}}{z_s} + \frac{1}{1+k_*^2/(a^2 H^2)} \right ] \bigg \} _\mathrm{pin}, 
\end{align}
where $\pi_{sk}$ and $\sigma_{sk}$ are the real and the imaginary part of $\psi_{sk}$ respectively. We set the initial phase of the mode function $u_{sk}^{(0)}$ to be zero for simplicity, which can be absorbed into the initial phase of the vacuum.
%%%%%

%subsection-----------------------------------------------------------------------------------------------------
\subsection{Slow roll expansion vs the large $N_*$ expansion}
%%%%%%
Using the slow roll expansion, the spectral index $n_s$ for the Starobinsky model is derived as Eq.~(\ref{nsv}), and its numerical value gives
\begin{align}
\label{ns1}
&n_s(k_*) \nonumber \\
 &\simeq \left [ 1-2 \epsilon _1 -\epsilon _2 -2 \epsilon _1 ^2 -(2C+3) \epsilon _1 \epsilon _2 - C \epsilon _2 \epsilon _3 \right ] _{n=n_*} \nonumber \\
& \simeq 0.9643\ .
\end{align}
%where $C=-0.729637$. 
Whereas, the large $N_*$ expansion up to $O(1/N_*^2)$ for Eq.~(\ref{nsv}),  in which the slow roll parameters are expanded by $1/N_*$, gives the following expression \cite{Martin:2016iqo}: 
\begin{align}
\label{N}
n_s(k_*) &\simeq 1-\frac{2}{N_*} + \left (0.81 + \frac{3}{2} \ln (N_*)\right )  \frac{1}{N_* ^2}  
                         \nonumber \\
& \simeq 0.9653. 
\end{align}
These two expansions differ slightly within the range of $1-2 \permil$. 
Since the large $N_*$ expansion here is just an approximation up to $O(1/N_*^2)$ and the definition of $N_*$ is entirely based on the slow roll approximation\footnote{
In the large $N_*$ expansion, 
the end of inflation is defined as the time when
\begin{equation}
\label{ }
\epsilon_v:= \frac{M_{\rm pl}^2}{2}\left(\frac{V_{,\phi}}{V}\right)^2 = 1 \ . \nonumber 
\end{equation}
While, the e-folding number from the end to the horizon crossing time is defined as
\begin{equation}
\label{ }
N_{*}^{\rm large}:= \frac{1}{M_{\rm pl}^2}\int^{\phi_*}_{\phi_{\rm end}}\frac{V}{V_{,\tilde{\phi}}}d\tilde \phi \ , \nonumber
\end{equation}
which does not include the $\dot \phi$ dependence of $N_*$. In the large $N_*$ expansion,
we expand the slow roll parameters defined in Eq. (\ref{srpara}) by the above $N_*^{\rm large}$ and calculate physical quantities like the spectral index. See \cite{Martin:2016iqo} for the details.}
, the result based on the slow roll expansion (\ref{ns1}) would be more reliable than that of the
large $N_*$ expansion (\ref{N}). 
If the sensitivity of future observations can be up to $O(10^{-3})$ for the scalar spectral index, it would be better to use numerical results of the slow roll expansion as the theoretical prediction for a fixed $N_*$.

%subsection-------------------------------------------------------------------------------------------------
\subsection{Analysis of the other expression for the spectral index}
%%%%%%%%%%%%%
%%%%%%%%%%%%%%
Here, we use the expression (\ref{npsi}) for the scalar perturbation
\begin{equation}
n_s = 2-4 \mathrm{Im} \left [ \frac{\psi _{sk}^{*2}(\tau_s)}{|\psi _{sk}(\tau_s) |^2} \int ^{\tau_s}_{-\infty(1+i\varepsilon)}  z_s ^2 (\partial_{\tilde{\tau}_s}\psi _{sk})^2 d \tilde{\tau}_s \right ] 
\label{nse1}
\end{equation}
to evaluate the scalar spectral index $n_s$.

First, as a schematic example, we consider the above expression for the perturbation on the exact de Sitter background. In that case, we obtain the exact solution for the mode function as Eq.~(\ref{u0}), and we give $z_s = a$. Using a new variable $x_s:= k\tau_s$ and Eq.~(\ref{psik}), the expression (\ref{nse1}) reduces to
\begin{align}
\label{nds}
    n_s^{\rm dS}= 2 - 4{\rm Im}&\left [ -\frac{1-x_s^2-2ix_s}{1+x_s^2}{\rm e}^{2ix_s}\right . \nonumber \\
             &\times \left.\int^{x_s}_{-\infty(1+i\varepsilon)}-\frac{1}{2}{\rm e}^{-2i\tilde{x}_s}d\tilde{x}_s \right ]   \ .
\end{align} 
Note that $x_s$ is always a negative number.
We can easily perform the integration and get the exact solution
\begin{equation}
\label{nex}
n_s^{\rm dS} = 1+ \frac{2x_s^2}{1+x_s^2}\ ,
\end{equation}
which is consistent to a direct calculation from Eq.~(\ref{u0}), (\ref{ps}) and (\ref{nso}). 
Then, we consider Eq. (\ref{nds}) for the time $|x_{s{\rm end}}|\ll 1$. Up to the second order of $x_{s{\rm end}}$, we obtain
\begin{align}
\label{nsin}
    &n_s^{\rm dS}|_{\rm end} \nonumber \\
    &\simeq 2 -4{\rm Im}\left[ -1\cdot -\frac{1}{2}\int^{x_{s\rm end}}_{-\infty(1+i\varepsilon)}
    {\rm e}^{-2i\tilde{x}_s} d\tilde{x}_s\right] \nonumber \\
%    &= 2+2\int^{x_{s\rm end}} {\rm sin}\,2\tilde{x}_s d\tilde{x}_s + O(x_{s\rm end}^3) \nonumber \\
    &= 2 - {\rm cos}\,2x_{s\rm end}  \nonumber \\
    &\simeq 1+ 2x_{s\rm end}^2  \ .
\end{align}
This is also consistent to the exact solution (\ref{nex}) up to the second order.
We find that only the imaginary part of the above integration contributes to the spectral index at the end $|x_{s\rm end}|\ll1$.
%The sine function is the imaginary part of the integrand in (\ref{nds}), and 
%We find that for $|x_{s\rm end}|\ll1$, the imaginary part of the integrand mostly contributes to the spectral index. 
For a further analysis, we separate the integration in (\ref{nsin}) as follows:
\begin{align}
\label{split}
    &{\rm Im}\left[\int^{x_{s\rm end}}_{-\infty(1+i\varepsilon)}{\rm e}^{-2i\tilde{x}_s}d\tilde{x}_s 
      \right] \nonumber \\
     =&{\rm Im}\left[\left(\int^{x_{s\rm end}}_{x_{s>}} + \int^{x_{s>}}_{x_{s\bigtriangleup}} + 
     \int^{x_{s\triangle}}_{-\infty(1+i\varepsilon)}\right){\rm e}^{-2i\tilde{x}_s}d\tilde{x}_s
      \right] \ ,
\end{align}
where $x_{s>}$ and $x_{s\triangle}$ satisfy
\begin{align}
\label{condi}
  1\gg |x_{s>}| \gg |x_{s\rm end}|,\quad {\rm sin}\,2x_{s\triangle} = \pm 1 \ .
\end{align}
%
%From Eq.~(\ref{nsin}), we find that only the imaginary part of the above integration contributes to the spectral index at the end $|x_{s\rm end}|\ll1$.
$x_{s\triangle}$ is the time when ${\rm sin}\,2x_s$ is at any of extrema, and thus the third part of integration becomes zero:
\begin{align}
\label{}
    &{\rm Im}\left[\int^{x_{s\triangle}}_{-\infty(1+i\varepsilon)}{\rm e}^{-2i\tilde{x}_s}d\tilde{x}_s 
       \right] \nonumber   \\
    =& {\rm Im}\left[ \frac{1}{2}({\rm sin}\,2\tilde{x}_s +i\,{\rm cos}\,2\tilde{x}_s )\right]^{x_{s\triangle}}_{-\infty(1+i\varepsilon)}  \nonumber  \\
    =& 0 \ .
\end{align}
This implies roughly that from the sub horizon scales, there is no contribution to the spectral index at the time $x_{s\rm end}$ since the integrand oscillates periodically. We take $x_{s\triangle}$ as the time corresponding to the last extremum of the sine function after the horizon crossing. 
%This matches to the fact that inside the horizon, the perturbation behaves as in Minkowski spacetime, and thus the spectral index is  
The first part of integration (\ref{split}) is the contribution to $n_s^{\rm dS}(x_{s\rm end})$ from the super-horizon scales
where $|x_{s}|\ll 1$, and the second part is that from only a few e-folds around the horizon crossing. 
Up to the second order of $x_{s>}$, the contribution from the first part becomes
\begin{align}
\label{}
    {\rm Im}\left[ \int^{x_{s\rm end}}_{x_{s>}}{\rm e}^{-2i\tilde{x}_s}d\tilde{x}_s\right] 
    &= {\rm Im}\left[\int^{x_{s\rm end}}_{x_{s>}}-i\,{\rm sin}\,2\tilde{x}_sd\tilde{x}_s\right]
     \nonumber \\
    &=\frac{1}{2}[{\rm cos}\,2x_{s\rm end}-{\rm cos}\,2x_{s>}]  \nonumber \\
    &\simeq x_{s>}^2\ ,  
\end{align}
where we ignore $O(x_{s\rm end}^2)$ terms. This shows clearly that there is less contributions 
to $n_s^{\rm dS}(x_{s\rm end})$ from the super-horizon scales, which matches
to the conservation of the perturbation (see Fig.~1).
In the following, we ignore the contribution from the super-horizon scales.
Eventually, %only the second part practically contributes to $n_s^{\rm dS}(x_{s\rm end})$, and
in Eq.~(\ref{nsin}), we can reduce the integration region to around the horizon crossing time only and extract the imaginary part of the integrand:  
\begin{align}
\label{nn}
    n_s^{\rm dS}|_{\rm end}&\simeq 2 -4{\rm Im}\left[ -1\cdot \frac{i}{2}\int^{x_{s>}}_{x_{s\bigtriangleup}}{\rm sin}\,2\tilde{x}_sd\tilde{x}_s \right]
       \nonumber   \\
    &=1 + O(x_{s>}^2) \ .
\end{align}
The accuracy of integration is up to $O(x_{s>}^2)$. 

\begin{figure}[htbp]
\begin{center}
\includegraphics[clip, bb = 0 0 720 540, width=1.0\columnwidth]{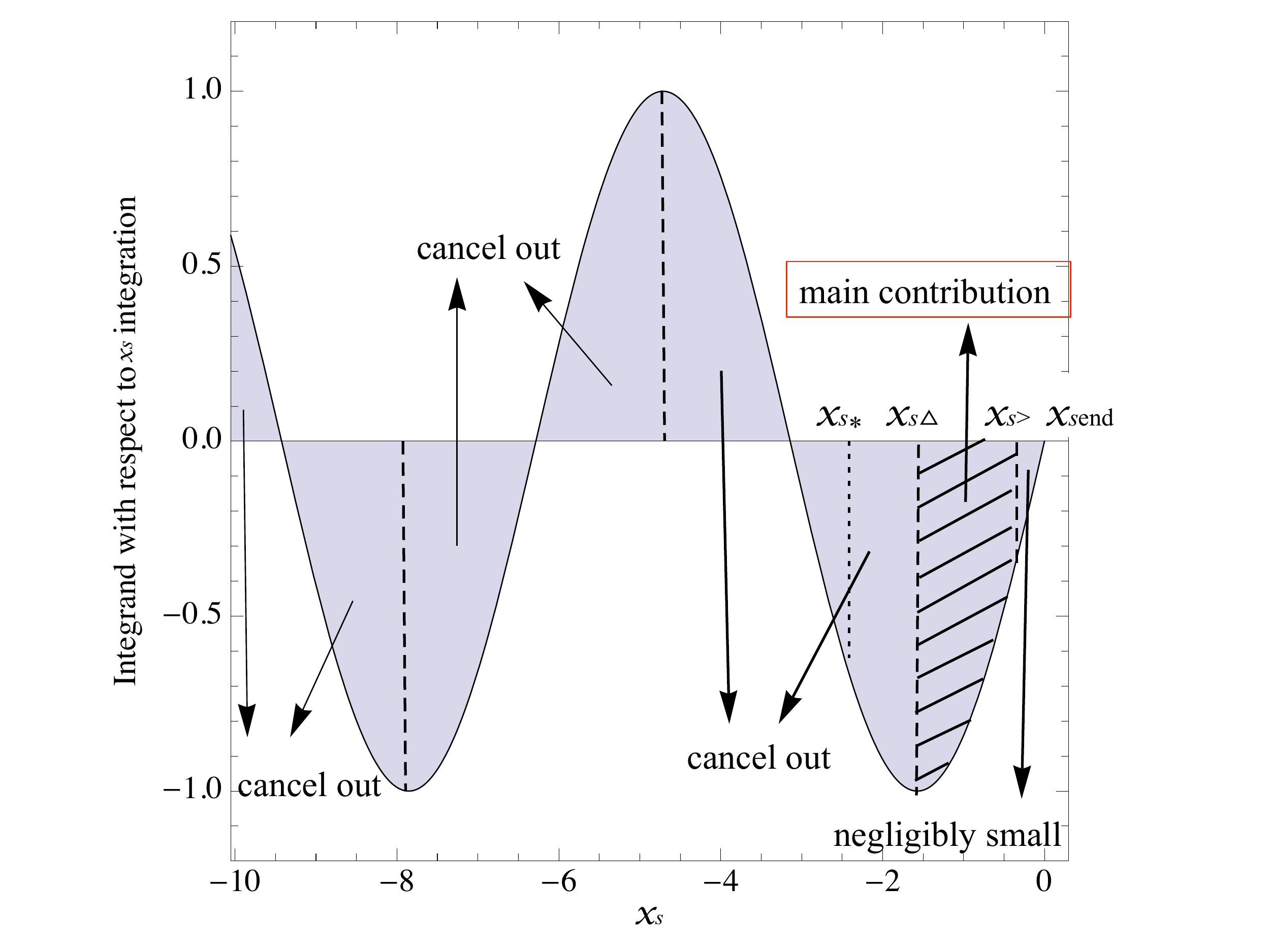}
\end{center}
\caption{Schematic picture of the integration in Eq.~(\ref{nsin}). Main contribution of the integration comes from the 
last quarter period of the imaginary part of integrand since the integrand with respect to $x_s$ is  
expressed by the sine function. }
\label{fig1}
\end{figure}

We expect a similar behavior of integrand for the generic case also. The situation, however,
changes importantly because the slow roll parameters will deform the integrand from the exact de Sitter case. We can understand the effect of slow roll parameters by looking at the MS equation (\ref{MS}) for the generic cases
\begin{align}
\label{}
   &\partial_{\tau_s}^2u_{sk} + \left[ k^2 %-\frac{\partial_{\tau_s}^2z_s}{z_s}
   + m_s^2(\tau_s)\right]u_{sk}=0 \ , \nonumber \\
   &m_s^2= -\frac{\partial_{\tau_s}^2z_s}{z_s} \nonumber \\
   &=  -\frac{a^2H^2}{c_s^2}\left\{ 2-\epsilon_1 + \frac{1}{4}(f_{s1}+5g_{s1})\right. \nonumber \\
   &- \frac{1}{4}\bigg[ (f_{s1}+g_{s1})\left(\epsilon_1 + \frac{1}{4}(f_{s1}-3g_{s1})\right) \nonumber \\
    &-f_{s1}f_{s2} - g_{s1}g_{s2}\bigg] \Bigg\} \ .
\end{align}
In the sub horizon scales where $c_sk\gg aH$, the oscillation term dominates over the mass term if the slow roll parameters remain not significantly big. In such a region, the mode function does not feel its mass during one period. Thus, as in Fig.~\ref{fig1}, the cancellation by the oscillation takes place in the deep sub horizon scales even if the slow roll parameters deform the mass. While, in the super-horizon scales where $c_sk\ll aH$, the mass term dominates over the oscillation term
and the mode function $u_{sk}$ rapidly approaches to the asymptotic value (\ref{asy}). 
%, and we restrict only to the case where $u_{sk}$ approaches rapidly to $A_s(k)z_s$. This corresponds that $\psi_{sk,n}$ approaches to zero because $\psi_{sk}=u_{sk}/z_s$ becomes almost a constant. 
Even if the slow roll expansion breaks down at the region $c_sk\ll aH$, the asymptotic behavior of mode function pursues as long as the mass term dominates enough over the oscillation term. 
Thus, there is less contribution from the super-horizon scales also even for the generic cases. The breakdown of slow roll expansion on the super-horizon scales implies a deformation of just the tiny fraction between $x_{s>}$ and $x_{s\rm end}$ in Fig.~\ref{fig1}, which remains still negligible. On the other hand, the mode function can be significantly deformed around the horizon crossing time, since it can feel the deformation of mass from the exact de Sitter case during a few e-folds around the horizon crossing time. Therefore, unlike the de Sitter case, we need to choose $x_{s\triangle}$ as the time when the imaginary part of integrand (\ref{nse1}) is at the extremum a few e-folds \textit{before} the horizon crossing time for the generic cases.   

%%%%%%%%%
%%%%%%%%%%

\begin{figure}[htbp]
\begin{center}
\includegraphics[clip, bb= 0 0 259 234, width=0.75\columnwidth]{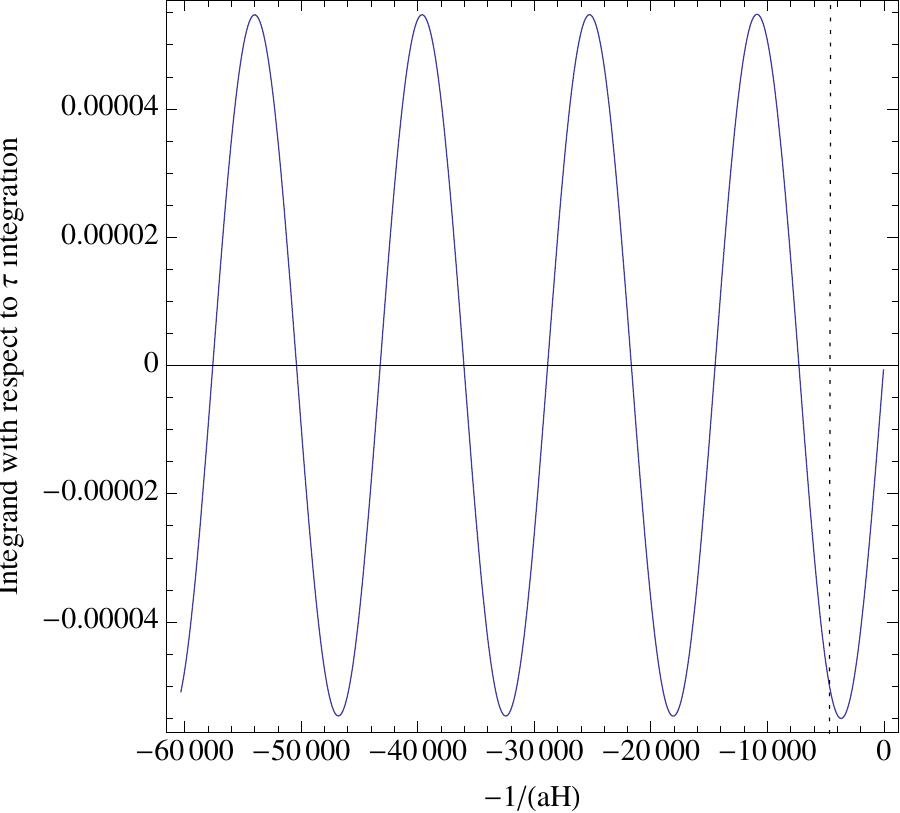}
\end{center}
\caption{The integrand in Eq.~(\ref{nsnf}) with respect to $-1/(aH) \simeq \tau_s$. We multiplied $aH$ to the integrand since here, we change the measure as $dn = aHd\tau_s$.
Similarly to the de Sitter case, the integrand (multiplied $aH$) almost behaves as the sine function.} 
\label{fig2}
\end{figure}

\begin{figure}[htbp]
\begin{center}
\includegraphics[clip, bb= 0 0 259 259, width=0.75\columnwidth]{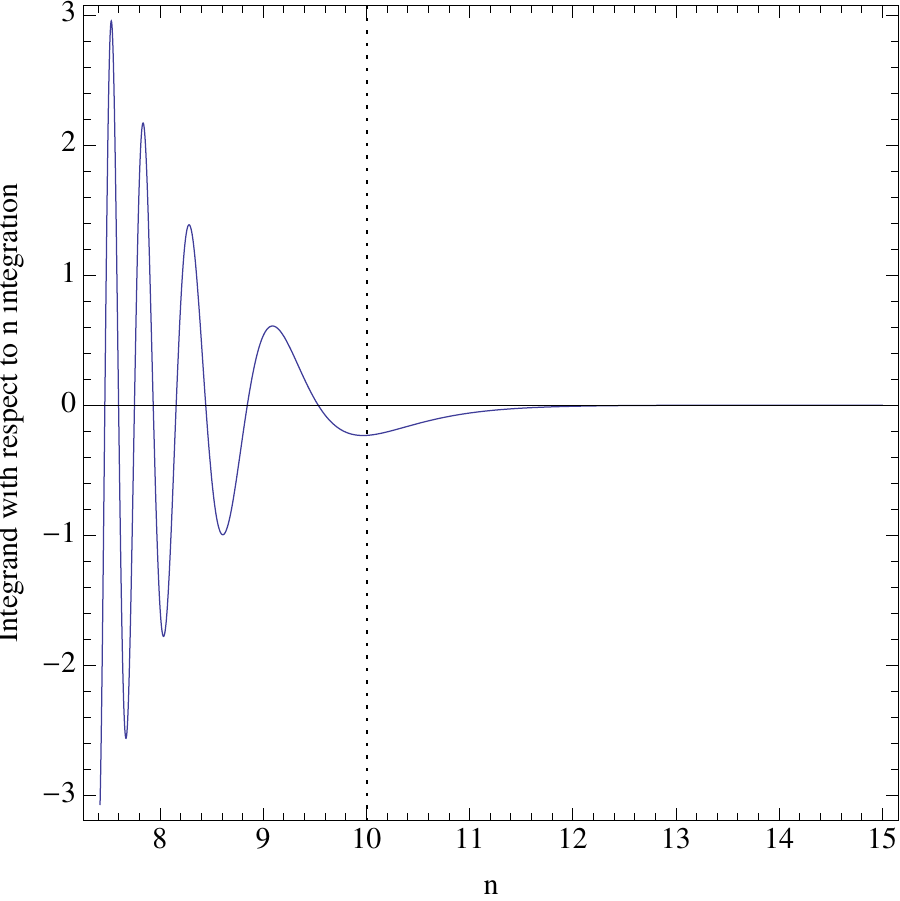}
\end{center}
\caption{The integrand in Eq.~(\ref{nsnf}) with respect to $n$. The last extremum appears after
the horizon crossing time $n_*=10$. }
\label{fig3}
\end{figure}

Following the speclation in the above, we reduce the integration region 
of Eq.~(\ref{nse1}) in terms of $n$ as 
\begin{align}
n_s|_{\rm end} %&\simeq 2-4 \mathrm{Im} \left [ \frac{\psi _{sk}^{*2}(\tau_{s\rm end})}{|\psi _{sk} (\tau_{s\rm end})|^2} \int ^{\tau_{s>}} _{\tau_{s\triangle}} z_s ^2 (\partial_{\tilde{\tau}_s}\psi _{sk})^2 d \tilde{\tau}_s \right ]  \nonumber \\
                 \simeq 2-4 \mathrm{Im} \left [ \frac{\psi _{sk}^{*2}(n_{\rm end})}{|\psi _{sk} (n_{\rm end})|^2} \int ^{n_>} _{n_\triangle} z_s ^2 aH \psi _{sk,\tilde n}^2 d \tilde n \right ] ,
\label{nse2}
\end{align}
where $n_>$ is a time a few e-folds after the horizon crossing and $n_\triangle$ is a time when the imaginary part of integrand is at the extremum.
With use of the decomposition $\psi_{sk}= \pi_{sk} + i \sigma_{sk}$, we write the prefactor of integration as
\begin{align}
\label{}
    \frac{\psi _{sk}^{*2}}{|\psi _{sk}|^2}\bigg|_{\rm end}= \frac{\pi_{sk}^2 -\sigma_{sk}^2- 2i\pi_{sk}\sigma_{sk}}{\pi_{sk}^2+\sigma_{sk}^2}\bigg|_{\rm end} \ .
\end{align}
In the de Sitter example, we can safely ignore the imaginary part of the prefactor since $\pi_{sk}$ for the de Sitter case approaches to zero on the super-horizon scales.  However, 
it is not for the present case because we changed the argument between $\pi_{sk}$ and $\sigma_{sk}$ to perform the numerical calculation.
Thus, to apply the analysis obtained from the de Sitter case for the present case, we need to tune the phase of $\psi_{sk}$ so that the real part of $\psi_{sk}(n_{\rm end})$ vanishes. 
%\footnote{Strictly speaking, the real part of $\psi_{sk}$ at the end of inflation does not vanish completely. From the analogy of the de Sitter case, however, we can expect the ratio of the real part to the imaginary part to be $\sim k_*/(a_{\rm end}H_{\rm end})\sim O({\rm e}^{-50})$. Thus, even if we tune the real part of $\psi_{sk}(n_{\rm end})$ exactly equal to zero, it hardly affects to the final result.}. 
We can always perform this procedure without a loss of generality by adding the phase factor to $\psi_{sk}$ as
\begin{align}
\label{}
    \psi_{sk}^{{\vartheta}}&= \pi_{sk}^{\vartheta} + i\sigma_{sk}^{\vartheta} 
                    := \psi_{sk}{\rm e}^{i{\vartheta}} \ , \nonumber \\
    {\rm cos}{\vartheta} &= \frac{\sigma_{sk}}{\sqrt{\pi_{sk}^2+\sigma_{sk}^2}}\bigg|_{\rm end},
    \quad {\rm sin}{\vartheta}= \frac{\pi_{sk}}{\sqrt{\pi_{sk}^2+\sigma_{sk}^2}}\bigg|_{\rm end}  \ ,              
\end{align}
which leads to $ \pi_{sk}^{{\vartheta}}(n_{\rm end}) = 0$. 
%Note that the phase ${\vartheta}$ is a constant number.
Then, using $\psi_{sk}^{\vartheta}$, we can rewrite Eq.~(\ref{nse2}) as
\begin{align}
\label{nsnf}
    n_s|_{\rm end}%&\simeq 2 - 4{\rm Im}\left[ -1\cdot \int ^{\tau_{s>}} _{\tau_{s\triangle}} 2z_s ^2
    %\partial_{\tilde{\tau}_s}\pi_{sk}^{\vartheta} \partial_{\tilde{\tau}_s}\sigma_{sk}^{\vartheta}
    %d \tilde{\tau}_s \right ] \nonumber   \\
   \simeq 2 - 4{\rm Im}\left[ -1\cdot i\int ^{n_>} _{n_\triangle}2\, z_s ^2 aH 
           (\pi_{sk,\tilde n}^{\vartheta} \sigma_{sk,\tilde n}^{\vartheta}) \,d \tilde n \right ] \ , 
\end{align}
where 
\begin{align}
\label{}
    &\pi_{sk, n}^{\vartheta} \sigma_{sk,n}^{\vartheta} 
      = (\pi_{sk,n}^2-\sigma_{sk,n}^2){\rm cos}{\vartheta}\,{\rm sin}{\vartheta}  \nonumber \\
        &\qquad\qquad\ + \pi_{sk,n}\sigma_{sk,n} ({\rm cos}^2{\vartheta} - {\rm sin}^2{\vartheta}) \ .
\end{align}
%
%If we decompose $\psi _{sk}$ by $\psi _{sk} = \pi + i \sigma $, we obtain 
%\begin{align}
%n_s -1 \simeq 1 &-4 \frac{2}{\pi ^2 + \sigma ^2} 
%\bigg [ (\pi ^2 - \sigma ^2) \int^{n_>}_{n_<} z_s^2 aH \pi_{,n} \sigma_{,n} d \tilde{n} \nonumber \\
%&- \pi \sigma \int^{n_>}_{n_<} z_s ^2 aH (\pi_{,n}^2 -\sigma_{,n} ^2) d \tilde{n} \bigg ] . 
%\end{align}
%By choosing $\pi$ and $\sigma$ as $\pi _\mathrm{end}$ and $\sigma _\mathrm{end}$, 
%we have the final expression of $n_s$ for the numerical calculation: 
%\begin{align}
%n_s -1  \simeq & 1 -8 \int ^{n_>} _{n_<} z_s^2 aH I_0 d \tilde{n} , \nonumber \\
%I_0 := & ( \pi _{,n}^2 -\sigma _{,n}^2 ) \frac{\pi _\mathrm{end} \sigma _\mathrm{end}}{\pi _\mathrm{end}^2 + \sigma _\mathrm{end}^2}  \nonumber \\
%&+ \pi _{, n} \sigma _{,n} \frac{\sigma _\mathrm{end}^2 - \pi _\mathrm{end}^2 }{\pi _\mathrm{end}^2 + \sigma _\mathrm{end}^2} . 
%\label{nsnf}
%\end{align}
%\RS{このへんに、πとσの位相を調節したことを明記。}
%
%The imaginary part of integrand with respect to $\tau$ approximately behaves as the sine-wave (see Fig.~\ref{fig2}). 
The actual behaviors of the integrand in Eq.~(\ref{nsnf}) are shown in Figs.~\ref{fig2}, \ref{fig3}. 
%They behave similarly as the de Sitter case as expected.   
%%%%%%%%%
%%%%%%%%%
%%%%%%%%
%%%%%%%%%%

As shown in Fig.~\ref{fig3}, the last extremum of the integrand in terms of $n$ is at $n=10.22055$, which is the e-folds soon after the horizon crossing time $n_*=10$. 
Then, we choose $n_\triangle$ as the time at the extremum before $n_\triangle = 10.22055$ and evaluate Eq.~(\ref{nsnf}) by choosing enough large value for $n_>$. 
The results are shown in Table \ref{tab1}. 
%%%%%%
\begin{table}[htp]
\caption{\label{tab1}Values of $n_s|_{\rm end}$ with respect to the position of $n_\triangle$}
\begin{tabular}{cccr}
\hline\hline
$i$-th last extremum &$n_\triangle$ &$ n_>$ &$ n_s|_{\rm end}$\\
\hline
1&10.22055&15&0.9759\\
2&9.13269&15&0.9632 \\
3&8.62547&15&0.9648 \\
4&8.29068&15&0.9644 \\
5&8.04033&15&0.9646\\
6&7.84027&15&0.9645\\
7&7.67365&15&0.9646\\
8&7.53086&15&0.9646 \\
\hline\hline
\end{tabular}
\end{table}
%%%%%%%%%%%
%
It shows that the deformation of the integrand from the exact de Sitter case causes
the red spectral tilt, and $n_s|_{\rm end}$ seems to converge into the value $0.9646$. This result is consistent with both of the slow roll expansion method and the large $N_*$ expansion method within the range of $1 \permil$ differences. 
The slight difference would originate from the approximations for the integration region and for the initial values of the mode function (\ref{pini}).

As we expected, the numerical value of spectral index from Eq.~(\ref{nsnf}) almost corresponds to that of horizon crossing formalism based on the slow roll expansion.
% by counting the contribution around from the horizon crossing time only.  
Thus, the analysis we performed for the integration region is legitimate. 
To calculate the spectral index, it is the most important to know the behaviors of the mode function and the slow roll parameters around the horizon crossing time. In turn, even if the slow roll expansion breaks down around the horizon crossing time only, we can no longer apply the horizon crossing formalism to the spectral index. This is because the mass around the crossing will be strongly deformed from the de Sitter case. %\RE{This interpretation is consistent with the well-known result of the Starobinsky model with the linear scalar potential \cite{Starobinsky:1992ts}.他の例ごと丸ごとぶち込んどくべきか？} 
We note that even if all the slow roll parameters are smaller than $O(1)$ values, the slow roll expansion can break when we cannot use the slow roll parameters as the expansion parameters.
%, $1>O(\epsilon_{i}^2)\gtrsim O(\epsilon_i)$, for example. 
Hence, if the potential has a bump and/or a cliff and the slow roll expansion breaks there, we need to perform the numerical analysis for the spectral index as here instead of using the horizon crossing formalism. We can apply the same analysis to the tensor perturbation also.

%section5===============================================================
\section{Summary}\label{sec5}

In this article, we have directly derived the two exact expressions of the spectral indices from the Ward-Takahashi identity associated with the dilatation charge.  
We use generic free actions for the scalar and the tensor perturbations, so that %The derived expressions (\ref{ns})-(\ref{npsi}) are generic, therefore, 
we can apply the expressions to all of the single field inflation models which possess the Lorentz-invariant scaling. 
In the slow roll expansion, we have used one of the expressions (\ref{ns}) and have derived the spectral indices up to the second order of the slow roll parameters. Applying our expression to the canonical scalar model with the potential, the result is consistent with the previous works.
Our formalism is, however, more rigorous and generic since we do not need any ansatzes nor approximations to the models and is therefore applicable to much more models than before.  
We have revealed that in the slow roll expansion, to evaluate the final values of the spectral indices at the end of inflation, we just need the horizon crossing formalism in which we change the time in the final values to the time when $a_*H_*=k_*$,
since the final values do not depend on the time actually.
%and thus we can evaluate them at an arbitrary time.  
%Therefore, the expressions (\ref{ns})-(\ref{npsi}) can be regarded as the generalization of well-known expressions of $n_s$ derived by horizon crossing formalism.  
We have also derived the runnings of the spectral indices up to the third order. 

To understand how the perturbations and the background contribute to the spectral indices, we have analyzed the other expression of the spectral indices (\ref{npsi}).
%The new expressions of the spectral indices  (\ref{ns})-(\ref{npsi})
First, we analyzed the scalar perturbation on the exact de Sitter background as a schematic example. Then, we applied the same method to the Starobinsky model and performed the numerical calculation. 
We found that the contributions to the spectral indices are only from around the horizon crossing time, and there is less contribution from the sub and the super-horizon scales. 
The spectral tilts are intrinsically connected to the masses of perturbations, deformed by the slow roll parameters, around the horizon crossing time.
This is because only around that time, the perturbations can feel the deformation of masses from the exact de Sitter case. In turn, this result indicates that even if the slow roll expansion breaks down around the horizon crossing time only, we can no longer apply the slow roll expansion since the mode functions around the horizon crossing time significantly deviate from those of the slow roll expansion. 
%This interpretation matches nicely to the well known result of another Starobinsky model with a linear potential.
%horizon crossing formalism breaks down if we consider the inflation models which predict several horizon crossings, 
%in other words, the models have no accelerated expansion regime. 
In such a case, the approximate expression (\ref{nsnf}) we have derived will be a powerful tool to evaluate the spectral indices.  

In Sec.~\ref{sec4}, we have evaluated the values of the scalar spectral index in 
several methods. It was shown that there is $1 \permil$ discrepancy between horizon crossing formalism and the large $N_*$ expansion in the Starobinsky model. 
In fact, the large $N_*$ expansion only has $1 \permil$ level precision, therefore, this is an expected result. On the other hand, 
our method is consistent with both the horizon crossing formalism and the large $N_*$ expansion in $1 \permil$ accuracy. 
%Our method loose its preciseness because of inaccurate initial condition; the solution $u$ in de Sitter background.  
Therefore, our method may be regarded as a useful indicator of $n_s$, similar to that using the  large $N_*$ expansion.

%将来的にはマルチでもいけると期待できる。また、ほかの対称性、例えば特殊共形変換など、からも非自明な恒等式が得られると期待される。それらはfuture workで。
%作用の形がプロテクトできてないと、量子効果やらなんチャラでSR expansionが途中のみbreak downするようなことが自然と起きる可能性あり。だから、正しいnsの値を求めたかったら、たとえツリーレベルでさえ、クロス付近のSRパラの値に注意を払う必要があるよ。というのが我々からのメッセージの一つである。

%acknowledgement and reference============================================
\section*{Aknowledgement}

The authors thank Toshiya Namikawa and Misao Sasaki for fruitful discussions.
The authors' work is partially supported by Leung Center for Cosmology and Particle Astrophysics, National Taiwan Univ. (FI121).

%Appendix A=============================================================
\appendix
\section{Derivation of another form of the spectral index}\label{A}

First, we deform the integration included in $\Theta_I$ (\ref{Theta}) as follows;
\begin{align}
\label{u1}
    & \frac{u_{Ik}^{*2}}{|u_{Ik}|^2}\int^{\tau_I}_{-\infty}\theta_I u_{Ik}^2d\tilde{\tau}_I  \nonumber \\
    &= \frac{u_{Ik}^{*2}}{|u_{Ik}|^2}\int^{\tau_I}_{-\infty}\left( \frac{m_I^2}{2}u_{Ik}^2
    -\tilde{\tau}_I\partial_{\tilde{\tau}_I} u_{Ik}\cdot m_I^2u_{Ik}
    \right) d\tilde{\tau}_I \nonumber \\
    &\quad -\frac{u_{Ik}^{*2}}{|u_{Ik}|^2}\frac{\tau_{I0}}{2}m^2_{I0}u_{Ik0}^2 + {\rm real\ part} \ ,
\end{align}
where the quantities with subscript $0$ are evaluated at $\tau_{I0}=-\infty$. 
We do not show the real parts explicitly because they do not contribute to $\Theta_I$. 
The mass at the past infinity approaches to zero faster than $1/\tau_I$ if we extrapolate the slow roll parameters as they remain finite at the past infinity, so the second term in the right hand side becomes zero.
Then, using the equation of motion for $u_{Ik}$, we reduce Eq.~(\ref{u1}) to
\begin{align}
\label{u2}
    & \frac{u_{Ik}^{*2}}{|u_{Ik}|^2}\int^{\tau_I}_{-\infty}\theta_I u_{Ik}^2d\tilde{\tau}_I  \nonumber \\
    %&({\rm A1}) \nonumber \\
    &= \frac{u_{Ik}^{*2}}{|u_{Ik}|^2}\int^{\tau_I}_{-\infty}\left[ \frac{m_I^2}{2}u_{Ik}^2
    +\tilde{\tau}_I\partial_{\tilde{\tau}_I} u_{Ik} (\partial_{\tilde{\tau}_I}^2u_{Ik} + k^2u_{Ik})
    \right] d\tilde{\tau}_I \nonumber \\
    &\quad+ {\rm real\ part} \nonumber \\ 
    &= \frac{u_{Ik}^{*2}}{|u_{Ik}|^2}\int^{\tau_I}_{-\infty}\left[ \frac{m_I^2}{2}u_{Ik}^2
    -\frac{1}{2}(\partial_{\tilde{\tau}_I} u_{Ik})^2 - \frac{k^2}{2}u_{Ik}^2
    \right] d\tilde{\tau}_I \nonumber \\
    &\quad+ \frac{u_{Ik}^{*2}}{|u_{Ik}|^2}\left[ \frac{\tilde{\tau}_I}{2}((\partial_{\tilde{\tau}_I}u_{Ik})^2 + k^2u_{Ik}^2)\right]^{\tau_I}_{-\infty} + {\rm real\ part}\nonumber \\
    &= \frac{u_{Ik}^{*2}}{|u_{Ik}|^2}\int^{\tau_I}_{-\infty}\left[ \frac{m_I^2}{2}u_{Ik}^2
    -\frac{1}{2}(\partial_{\tilde{\tau}_I} u_{Ik})^2 - \frac{k^2}{2}u_{Ik}^2
    \right] d\tilde{\tau}_I \nonumber \\
    &\quad + \frac{\tau_I}{2}\frac{u_{Ik}^{*2}}{|u_{Ik}|^2}(\partial_{\tau_I}u_{Ik})^2 + {\rm real\ part} \ ,
\end{align}
where we used Eq.~(\ref{uasy}) at the last equality.
We can further deform this integration as
\begin{align}
\label{u3}
    & \frac{u_{Ik}^{*2}}{|u_{Ik}|^2}\int^{\tau_I}_{-\infty}\theta_I u_{Ik}^2d\tilde{\tau}_I \nonumber \\
    &= \frac{u_{Ik}^{*2}}{|u_{Ik}|^2}\int^{\tau_I}_{-\infty}\left[ -\frac{1}{2}(\partial_{\tilde{\tau}_I}^2u_{Ik}+ k^2u_{Ik})u_{Ik}
     \right. \nonumber \\
    &\qquad\qquad\quad\left. -\frac{1}{2}(\partial_{\tilde{\tau}_I} u_{Ik})^2 - \frac{k^2}{2}u_{Ik}^2
    \right] d\tilde{\tau}_I \nonumber \\
    &\quad+ \frac{\tau_I}{2}\frac{u_{Ik}^{*2}}{|u_{Ik}|^2}(\partial_{\tau_I}u_{Ik})^2 + {\rm real\ part} \nonumber \\
    &= - \frac{u_{Ik}^{*2}}{|u_{Ik}|^2}\int^{\tau_I}_{-\infty} k^2u_{Ik}^2
    d\tilde{\tau}_I - \frac{1}{2}u_{Ik}^*\partial_{\tau_I}u_{Ik} \nonumber \\
    &\quad+ \frac{1}{2}\frac{u_{Ik}^{*2}}{|u_{Ik}|^2}u_{Ik0}\partial_{\tau_I}u_{Ik0} +
    \frac{\tau_I}{2}\frac{u_{Ik}^{*2}}{|u_{Ik}|^2}(\partial_{\tau_I}u_{Ik})^2 \nonumber \\
     &\quad+ {\rm real\ part} \ .
\end{align}
Eliminating the $k^2$ term from Eq. (\ref{u2}) and (\ref{u3}), we obtain
\begin{align}
\label{u4}
    & \frac{u_{Ik}^{*2}}{|u_{Ik}|^2}\int^{\tau_I}_{-\infty}\theta_I u_{Ik}^2d\tilde{\tau}_I \nonumber \\
    &= \frac{u_{Ik}^{*2}}{|u_{Ik}|^2}\int^{\tau_I}_{-\infty}\left[ m_I^2u_{Ik}^2
    -(\partial_{\tilde{\tau}_I} u_{Ik})^2\right] d\tilde{\tau}_I \nonumber \\
    &\quad + \frac{1}{2}u_{Ik}^*\partial_{\tau_I}u_{Ik} + \frac{\tau_I}{2}\frac{u_{Ik}^{*2}}{|u_{Ik}|^2}(\partial_{\tau_I}u_{Ik})^2  \nonumber \\
    &\quad - \frac{1}{2}\frac{u_{Ik}^{*2}}{|u_{Ik}|^2}u_{Ik0}\partial_{\tau_I}u_{Ik0} 
    + {\rm real\ part} \ .
\end{align}
We take the following contour for the integration in Eq.~(\ref{u4}):
\begin{align}
\label{}
    &\int^{\tau_I}_{-\infty}\left( m_I^2u_{Ik}^2
    -(\partial_{\tilde{\tau}_I} u_{Ik})^2\right) d\tilde{\tau}_I \nonumber \\
    &= \left(\int^{\tau_I}_{-\infty(1+i\varepsilon)} + \int^{-\infty(1+i\varepsilon)}_{-\infty}\right)
    \left[ m_I^2u_{Ik}^2 -(\partial_{\tilde{\tau}_I} u_{Ik})^2\right] d\tilde{\tau}_I    \ ,
\end{align}
where we apply the $i\varepsilon$ prescription so as to calculate the above integration easily. We are not aware of the implication of this prescription. However, we do not need to take care about it since the extrapolated mode function before the inflation does not need to follow the actual history of the universe. It is enough to set the contour at the past infinity calculable.
We extrapolate the mode function at the second contour as the same form as Eq.~(\ref{uasy}). Then, the integration from the second contour gives
\begin{align}
\label{}
    &\int^{-\infty(1+i\varepsilon)}_{-\infty}
    \left[ m_I^2u_{Ik}^2 -(\partial_{\tilde{\tau}_I} u_{Ik})^2\right] d\tilde{\tau}_I   \nonumber \\
    = &\int^{-\infty(1+i\varepsilon)}_{-\infty}\frac{k}{2}{\rm e}^{-2ik\tilde{\tau}_I} d\tilde{\tau}_I \nonumber \\
    = &-\frac{i}{4}{\rm e}^{-2ik\tau_I}\bigg|_{\tau_I=-\infty} = \frac{1}{2}u_{Ik0}\partial_{\tau_I}u_{Ik0}\ .
\end{align}
Then, Eq.~(\ref{u4}) reduces to
\begin{align}
\label{u5}
    & \frac{u_{Ik}^{*2}}{|u_{Ik}|^2}\int^{\tau_I}_{-\infty}\theta_I u_{Ik}^2d\tilde{\tau}_I \nonumber \\
    &= \frac{u_{Ik}^{*2}}{|u_{Ik}|^2}\int^{\tau_I}_{-\infty(1+i\varepsilon)}\left[ m_I^2u_{Ik}^2
    -(\partial_{\tilde{\tau}_I} u_{Ik})^2\right] d\tilde{\tau}_I \nonumber \\
    &\quad + \frac{1}{2}u_{Ik}^*\partial_{\tau_I}u_{Ik} + \frac{\tau_I}{2}\frac{u_{Ik}^{*2}}{|u_{Ik}|^2}(\partial_{\tau_I}u_{Ik})^2  \nonumber \\
    &\quad + {\rm real\ part} \ .
\end{align}
Substituting this to Eq.~(\ref{Theta}), we get
\begin{align}
\label{}
    \Theta_I
    &= {\rm Im}\left[ 2u_{Ik}^*\partial_{\tau_I}u_{Ik} + \frac{2\tau_I}{|u_{Ik}|^2}u_{Ik}^{*2}(\partial_{\tau_I}u_{Ik})^2 \right. \nonumber  \\
    &\qquad\quad \left.+ \frac{4u_{Ik}^{*2}}{|u_{Ik}|^2}\int^{\tau_I}_{-\infty(1+i\varepsilon)}\left( m_I^2u_{Ik}^2
    -(\partial_{\tilde{\tau}_I} u_{Ik})^2\right) d\tilde{\tau}_I\right] \nonumber \\
    &= -1 -2\tau_I\frac{\partial_{\tau_I}|u_{Ik}|}{|u_{Ik}|} \nonumber \\
    &\quad + 4{\rm Im}\left[\frac{u_{Ik}^{*2}}{|u_{Ik}|^2}\int^{\tau_I}_{-\infty(1+i\varepsilon)}\left( m_I^2u_{Ik}^2
    -(\partial_{\tilde{\tau}_I} u_{Ik})^2\right) d\tilde{\tau}_I\right] \ ,
\end{align}
where we used the Wronskian (\ref{W}) at the second equality.
Introducing $\psi_{Ik}= u_{Ik}/z_I$ and performing a integration by parts, 
we obtain the spectral index in the other generic form (\ref{npsi})
\begin{align}
\label{}
    n_I&= {\cal N}_I + 1 - 4{\rm Im}\left[\frac{\psi_{Ik}^{*2}}{|\psi_{Ik}|^2}\int^{\tau_I}_{-\infty(1+i\varepsilon)}
    z_I^2(\partial_{\tilde{\tau}_I}\psi_{Ik})^2 d\tilde{\tau}_I\right] \ . \nonumber    
\end{align} 
%

%Appendix B========================================================
\section{Specific models for the slow roll expansion}\label{B}
\subsection{Power-law inflation}

We first consider the power-law inflation model as one of specific models. The scale factor and the coefficients in the free action  are given by
\begin{align}
\label{}
    a&\propto t^{1/\epsilon_1} \ ,  \nonumber \\
    {\cal F}_s&={\cal G}_s= \epsilon_1 = {\rm constant}\ , \nonumber \\
    {\cal F}_t&={\cal G}_t= 1 \ . 
\end{align}
In this case, for the scalar and the tensor perturbations, the conformal times and the masses  become the same ones
\begin{align}
\label{}
    \tau_s &= \tau_t= -\frac{1}{aH(1-\epsilon_1)} \ , \nonumber \\
    m_I^2&= -\frac{\partial_{\tau_I}^2a}{a} = -\frac{2-\epsilon_1}{\tau_I^2(1-\epsilon_1)^2}  \ .
\end{align}
We can find that the breaking size $\theta_I$ becomes exactly equal to zero
\begin{align}
\label{}
    \theta_I&= m_I^2 + \frac{\tau_I}{2}\partial_{\tau_I}m_I^2 = 0  \ .
\end{align}
Thus, for the power-law inflation model, the free actions preserve the dilatation symmetry.
The same is true for the free scalar field on the exact de Sitter background.
The asymptotic behavior of the mode function is governed by the growing mode
\begin{align}
\label{}
    u_{Ik}&\rightarrow A_Iz_I + B_Iz_I\int^{\tau_I}\frac{d\tilde{\tau}_I}{z_I^2} \nonumber   \\
    &\rightarrow A_Iz_I \qquad {\rm when}\quad |k\tau_I|\rightarrow 0\ ,
\end{align}
since
\begin{equation}
\label{ }
\int^{\tau_I}\frac{d\tilde{\tau}_I}{z_I^2}\propto a^{-(3-\epsilon_1)}\ .
\end{equation}
Consequently, we can obtain the asymptotic values of the spectral indices for the power-law inflation as
\begin{align}
\label{}
    n_s-1|_{|k\tau_I|\rightarrow 0}&= n_t|_{|k\tau_I|\rightarrow 0} =-\frac{2\epsilon_1}{1-\epsilon_1} \ .
\end{align}
%which reflects the dilation symmetry of the system.

%usr---------------------------------------------------------------------------------------------------
\subsection{Ultra slow roll inflation}

As another specific case, we consider the ultra slow roll inflation model. For this model, we cannot take the comoving gauge at the asymptotic future $|k\tau_I|=0$, but before that time, we can still work on the comoving gauge. 
The coefficients in the free action are given by
\begin{align}
\label{}
    {\cal F}_s&={\cal G}_s= \epsilon_1\propto a^{-6}\ , \nonumber \\
    {\cal F}_t&={\cal G}_t= 1 \ . 
\end{align}
In this model, the slow roll parameter $\epsilon_1$ rapidly decays in $-6$ powers of the scale factor, and thus other slow roll parameters become
\begin{align}
\label{}
    &f_{s1}= g_{s1}= \epsilon_2 = -6\ , \nonumber   \\
    &{\rm others} = 0\ .  
\end{align} 
The mass of $u_{I}$ and the breaking sizes become
\begin{align}
\label{}
    m_s^2&= -a^2H^2(2+ 2\epsilon_1) \ , \nonumber   \\
    m_t^2&= -a^2H^2(2 -\epsilon_1) \ , \nonumber \\
    \theta_s&= -a^2H^2\left[8\epsilon_1+2\epsilon_1^2-(2-6\epsilon_1+2\epsilon_1^2)\sum_{i=1}^\infty\delta_{si}\right] \nonumber  \\
    &= O(a^{-4}) \ , \nonumber \\
    \theta_t&= -a^2H^2\left[-\epsilon_1+\epsilon_1^2-(2-2\epsilon_1^2)\sum_{i=1}^\infty\delta_{ti}\right] \nonumber  \\
    &= O(a^{-4}) \ .  
\end{align}
Note that the summation of $\delta_{Ii}$ starts from $i=1$, and all of $\delta_{Ii\geq 1}$ is proportional to $\epsilon_1$ for this model.
If we ignore all of the products proportional to $\epsilon_1$, we can omit the breaking term $\theta_I$ or $\Theta_I$. It is well known that for this model, the scalar mode function on the super horizon scales is dominated by the decaying mode 
\begin{align}
\label{}
    u_{sk}&\rightarrow B_sz_s\int^{\tau_s}\frac{d\tilde{\tau}_s}{z_s^2} \quad {\rm when }\quad 0<|k\tau_I|\ll 1 \ ,
\end{align}
since the decaying mode actually grows faster than the growing mode on the super-horizon scales
\begin{align}
\label{decay}
    \int^{\tau_s}\frac{d\tilde{\tau}_s}{z_s^2}&= \int^{n}\frac{d\tilde{n}}{a^3H{\cal G}_s}\simeq \frac{1}{3a^3H{\cal G}_s}\propto O(a^3) \ .
   % &\simeq -\frac{z_s}{3a^3H{\cal G}_s} + 2z_s\int^{n}\frac{d\tilde{n}}{a^3H{\cal G}_s} \nonumber \ , \\
    %z_s\int^{\tau_s}\frac{d\tilde{\tau}_s}{z_s^2}&\simeq \frac{z_s}{3a^3H{\cal G}_s} \ .
\end{align}
The tensor mode function is dominated by the growing mode as usual. Using Eq.~(\ref{ns}), we obtain the spectral indices on the super-horizon scales for the ultra slow roll inflation
as
\begin{align}
\label{}
    n_s-1|_{|k\tau_I|\ll 1}&\simeq  2 + 2\tau_s\left(\frac{\partial_{\tau_s}z_s}{z_s} + \frac{1}{z_s^2\int^{\tau_s}\frac{d\tilde{\tau}_s}{z_s^2}} \right)  \nonumber \\
    &\simeq 0  \ , \\
    n_t|_{|k\tau_I|\ll 1} &\simeq 2 + 2\tau_t\left(\frac{\partial_{\tau_t}z_t}{z_t}\right) \nonumber \\
    &\simeq 0  \ ,
\end{align} 
where we ignore $O(\epsilon_1)$ terms since they rapidly decay into zero. While the horizon crossing formalism in the literature cannot estimate the scalar spectral index for the ultra slow roll inflation, our expression can predict it correctly since our expression is exact and can be applied to the generic single field models.

%Reference=====================================================
\bibliographystyle{apsrmp}
%\bibliography{rmp-sample}

\end{document}